\documentclass[useAMS]{mn2e}

\usepackage{graphicx}	
\usepackage{amsmath}	
\usepackage{amssymb}	
\usepackage{multicol}
\usepackage{longtable}

\usepackage{bm}		
\usepackage{pdflscape}	
\usepackage[utf8]{inputenc}
\usepackage[T1]{fontenc}
\usepackage{ae,aecompl}

\title[GAIA eDR3 wide binary kinematics] {Internal kinematics of GAIA eDR3 wide binaries.} 

\author[X. Hernandez,  S. Cookson and R. A. M. Cort\'{e}s] {X. Hernandez$^{1}$, S. Cookson$^{2}$, and  R. A. M. Cort\'{e}s$^{1}$\\ 
$^{1}$Instituto de Astronom\'{\i}a, Universidad Nacional Aut\'{o}noma de M\'{e}xico,
  Apartado Postal 70--264 C.P. 04510 M\'exico D.F. M\'exico. \\
$^{2}$Crawley Astronomical Society, c/o Copthorne Prep Sch, Effingham Ln, Copthorne, Crawley RH10 3HR, UK.\\
}

\date{Released 30/07/2021}

\begin{document}

\label{firstpage}

\maketitle

\begin{abstract}
  Using the recent GAIA eDR3 catalogue we construct a sample of solar neighbourhood isolated wide binaries satisfying
  a series of strict signal-to-noise data cuts, exclusion of random association criteria and detailed colour-magnitude diagram
  selections, to minimise the presence of any kinematic contaminating effects having been discussed in the literature to date. Our
  final high-purity sample consists of 423 binary pairs within 130 pc of the sun and in all cases high-quality GAIA single-stellar
  fits for both components of each binary (final average RUWE values of 0.99), both also restricted to the cleanest region of the
  main sequence. We find kinematics fully consistent with Newtonian expectations for separations, $s$, below 0.009 pc, with relative
  velocities scaling with $\Delta V \propto s^{-1/2}$ and a total binary mass, $M_{b}$, velocity scaling consistent with $\Delta V
  \propto M_{b}^{1/2}$. For the separation region of $s> 0.009$ pc we obtain significantly different results, with a separation independent
  $\Delta V \approx 0.5$ km/s and a $\Delta V \propto M_{b}^{0.24 \pm 0.21}$. This situation is reminiscent of the low acceleration
  galactic baryonic Tully-Fisher phenomenology, and indeed, the change from the two regimes we find closely corresponds to the $a \lesssim
  a_{0}$ transition. {  These results are at odds not only with Newtonian expectations, but also with MOND predictions, where the presence
  of an external field effect implies only small deviations from Newtonian dynamics are expected for Solar Neighbourhood wide binaries.}

\end{abstract}


\section{Introduction}

The most salient features of the gravitational anomalies generally ascribed to the presence of a dominant dark matter component
at galactic scales, are the loss of a dependence on radius for equilibrium velocities, the clear scaling of these equilibrium
velocities with the fourth root of the total baryonic mass of the systems in question, and the occurrence of the transition
from a regime where the observed matter alone adequately explains kinematic observations through Newtonian gravity, to the
aforementioned anomalous region, always at acceleration scales of $a=a_{0} \approx 1.2 \times 10 ^{-10}$ m/s$^{2}$.

The first two of the above
properties are of course the well known flatness and baryonic Tully-Fisher relation of late type galactic rotation curves, which can
be summarised as $V_{TF}=0.35 (M/M_{\odot})^{1/4}$ km/s (McGaugh et al. 2000), and which have been recently extended to the asymptotically
flat velocity dispersion profiles of a variety of pressure supported systems such as globular clusters (e.g. Scarpa et al. 2003,
Hernandez et al. 2012b, Hernandez \& Lara-D I 2020) and elliptical galaxies (e.g. Jimenez et al. 2013, Durazo et al. 2018, Chae et
al. 2020a), also showing a scaling with the fourth root of the total baryonic mass.

The last of
the three traits described above forms the basis of MOND (Milgrom 1983) as an alternative to the standard dark matter postulate,
where an underlying transition in the structure of physics is proposed as the causal mechanism behind the gravitational anomalies
which occur in the low acceleration regime at galactic scales, and forms also the central tenant of a large range of modified gravity
theories inspired by MOND, aiming at an explanation in the absence of the dark matter hypothesis. Examples of these last are
Bekenstein (2004), Moffat \& Toth (2008), Zhao \& Famaey (2010), Capozziello \& De Laurentis (2011), Verlinde (2016),
Barrientos \& Mendoza (2018), McCulloch et al. (2019) and Hernandez et al. (2019b).

Within the context described above, one of us in Hernandez et al. (2012a) identified wide binaries as an interesting test case
to explore the generality (or lack thereof) of the gravitational anomalies detected at galactic scales, to an entirely different
astronomical range of scales and masses, which however, shares the same low acceleration regime. Wide binaries composed of two
$1 M_{\odot}$ mass stars will cross the $a<a_{0}$ threshold when the separation between both components, $s$, becomes larger than about
7,000 AU, 0.035 pc, Hernandez et al. (2012a). That initial exploration of the problem was performed using the best astrometry
available at the time, that from the Hipparcos satellite. The results were for 1D relative velocities on the plane of the sky for a
small carefully selected sample of 280 wide binaries, showing values much larger than Newtonian expectations, and, to within the large
confidence intervals of that study, consistent with no dependence on the separation on the plane of the sky of the two components of
the binaries studied, for separations well within the tidal radius of the problem, of about $0.7$ pc (e.g. Jiang \& Tremaine 2010).

Since, the interest on the proposed test has grown and a number of independent investigations have been performed, confirming the
presence of the signal first detected in Hernandez et al. (2012a). Scarpa et al. (2017) performed detailed ground based
follow up observations of a small sub-sample of the stars analysed in Hernandez et al. (2012a), obtaining accurate spectroscopic
line of sight velocities, and confirming the presence of relative 3D velocities in excess of Newtonian expectations. Pittordis \&
Sutherland (2018), Banik \& Zhao (2018), and Banik \& Kroupa (2019) refined the original test and conclude that
distinguishing between Newtonian and modified gravity models will be possible using final GAIA data, once the full mission
accuracy becomes available. Using GAIA DR2 data Pittordis \& Sutherland (2019) show that some MOND variants appear disfavoured
by their analysis of wide binaries, while the uncertainties inherent to the data used were unable to unequivocally discern
between a purely Newtonian scenario and some MOND variants. More recently, in Hernandez et al. (2019a) two of us revisited the wide
binary sample of the original experiment, but using the much superior GAIA DR2 astrometry, and obtaining results consistent with the
original, in spite of substantially reduced confidence intervals on the relative velocities.


For relatively bright unresolved stellar companions, the position of the observed star in the HR diagram will
shift to brighter magnitudes as per standard spectroscopic binaries, while for sufficiently dim undetected companions (essentially
stellar dark matter), the wobble induced on the affected component of the observed binary, a kinematic contamination inflating
the detected binary kinematics,  will become apparent through the quality of the fit of the single star solution. In more detail,
Belokurov et al. (2020) show that unresolved stellar companions in the GAIA DR2 catalogue result in poor single stellar fits
(RUWE values larger than 1.4, extending to 10 or sometimes even larger, the re-normalised unit weight error parameter of GAIA), and
identify regions of the solar neighbourhood GAIA HR diagram where unresolved stellar companions are prominent, and regions where
such companions are largely absent.

In this paper we use the recent GAIA eDR3 (Gaia Collaboration et al. 2016, Gaia Collaboration et al. 2021, Lindegren et al. 2021,
Brown et al. 2020) catalogue to construct a large sample of 5,844 nearby, locally isolated
wide binaries having high quality radial velocities, parallaxes, colours, magnitudes and proper motions, to construct a detailed picture
of their relative velocity on the plane of the sky, $\Delta V$. This, as a function of both {  2D projected separation}
and total binary mass,$M_{b}$, {  the latter as a diagnostic tool to help derive physical insight into the trends observed in
velocity as a function of binary separation}, with stellar mass inferred from main sequence accurate estimates through the absolute
G magnitude from Pittordis \& Sutherland (2019).

The initial sample is then carefully pruned to minimise all sources of kinematic contamination discussed
to date, a strict RUWE parameter upper
limit designed to exclude short period blended tertiaries, relative radial velocity cuts to exclude contamination of unbound stars
from the mean field local distribution, and a detailed HR diagram region selection to further exclude the presence of unresolved
stellar companions, restricted to the cleanest section of the main sequence, always for both companions of each {  nearby} binary.
Results are consistent with Newtonian expectations in both separation and mass relative velocity scalings for the low separations
corresponding to a high acceleration regime, but reminiscent of Tully-Fisher scalings for the low acceleration regime
of wide binary separations.

 \begin{figure}
 \vskip 0pt
 \hskip -5pt \includegraphics[height=7.0cm,width=8.5cm]{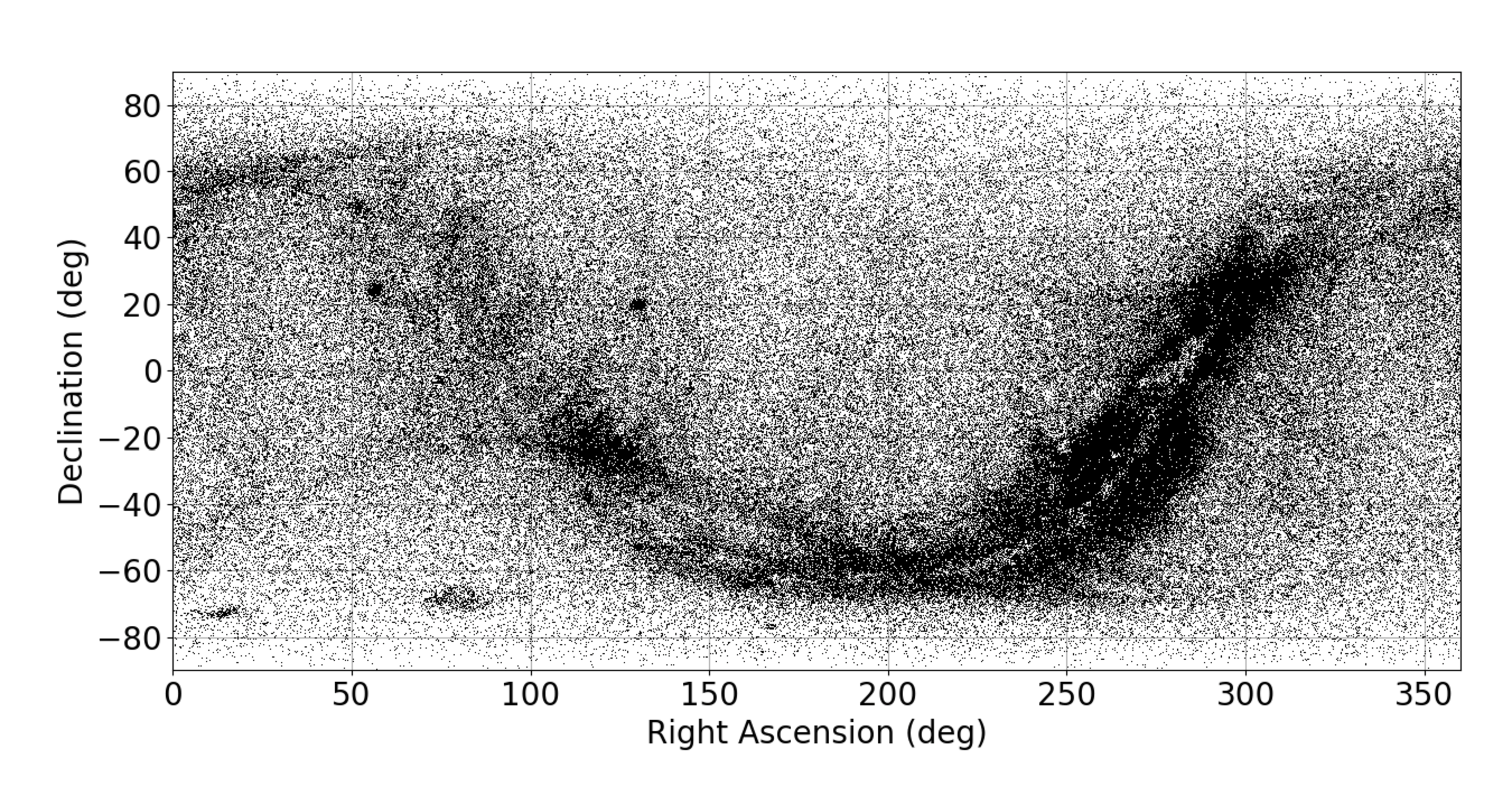}
 \caption{Full sky plot displaying over 800,000 binary pairs from the GAIA eDR3 catalogue satisfying our initial search criteria
   for binary star candidates within 200 pc of the Sun, see text. The presence of nearby groups and associations, as well as
 heavily crowded fields along the disk of the Milky Way are evident.}
 \end{figure}


In section 2 we present our initial sample selection describing the removal of binaries in clustered environments to produce
a catalogue of relatively isolated binaries, the use of a relative velocity along the line of sight threshold to exclude chance
associations of unbound systems, and the introduction of signal to noise quality cuts on the original GAIA eDR3 parameters.
Section 3 describes the cleaning of the sample by gradually introducing further distance cuts to minimise the probability of
keeping binaries with long period unseen companions, the use of a RUWE quality cut to exclude unresolved tertiaries, and the
implementation of a strict HR diagram selection criteria to isolate only binaries were both components are main sequence stars
with minimal probabilities of including blended tertiaries. {  In section 4 we compare our results to those of some similar
recent studies.} Finally, section 5 presents our concluding remarks.

\section{Initial Sample Selection}

We begin by using the GAIA initial search of El-Badry \& Rix (2018) who construct a catalogue of wide binaries within 200pc
of the sun, which
is then tested through various simulations to account for projection effects, a reasonable distribution of ellipticities and completion
due to undetected companions, with a level of contamination estimated by those authors as $< 0.2 \%$. An extension of the basic strategy
presented in that paper was also used in Tian et al. (2020) to construct a catalogue of over 800,000 wide binaries within 4 kpc of the
sun, maintaining a high purity for the sample.

\begin{figure}
 \vskip 0pt
 \hskip -5pt \includegraphics[height=7.0cm,width=8.5cm]{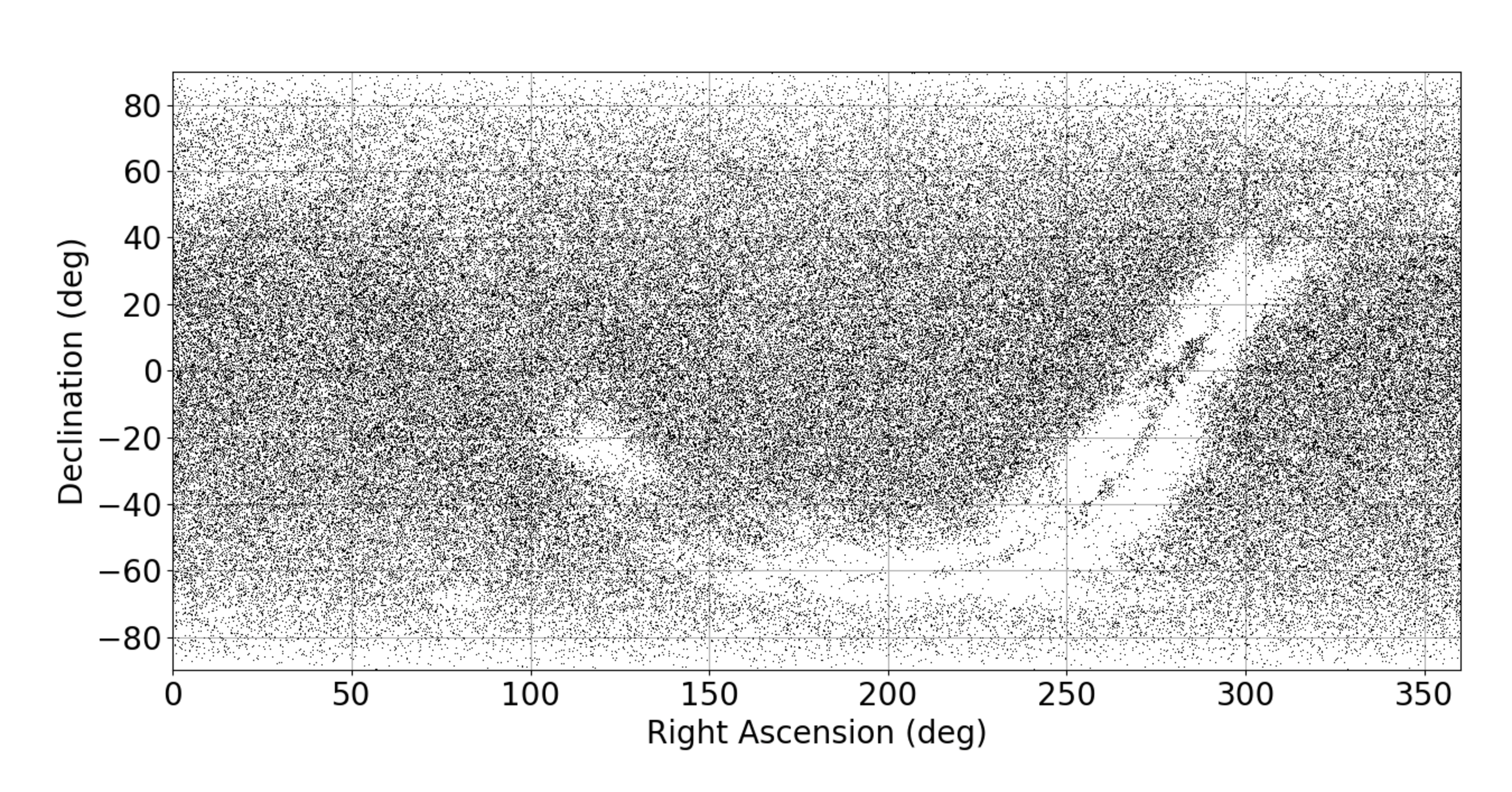}
 \caption{Full sky plot of 169,540 isolated wide binary pair candidates remaining after the removal of all candidate pairs
   were either the primary or the secondary star had been originally selected as either a primary or a secondary member of more
   than one binary pair candidate. The remaining binary pairs have no GAIA sources with parallax signal to noise determinations
   > 5 within a 3D sphere of 0.5 pc. The removal of all local groups, as well as large areas of crowded fields along the Milky Way disk
   projection is apparent.}
 \end{figure}

 \begin{figure}
 \vskip 0pt
 \hskip -5pt \includegraphics[height=7.0cm,width=8.5cm]{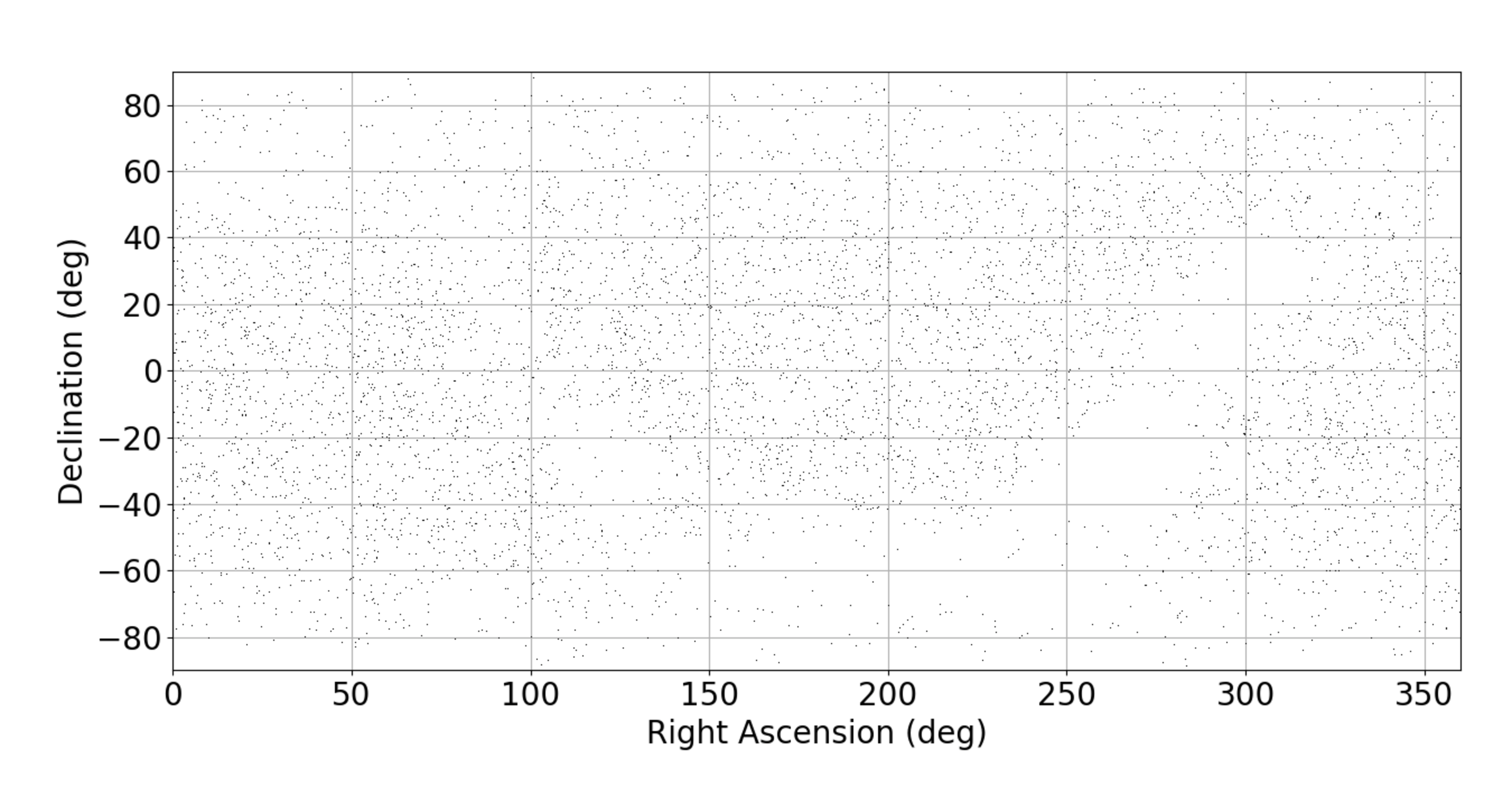}
 \caption{Full sky plot of all 5,844 binary pairs in figure 2 having radial velocity measurements for both primaries and
   secondaries.}
 \end{figure}

 \begin{figure*}
    \includegraphics[height=6.0cm,width=\columnwidth]{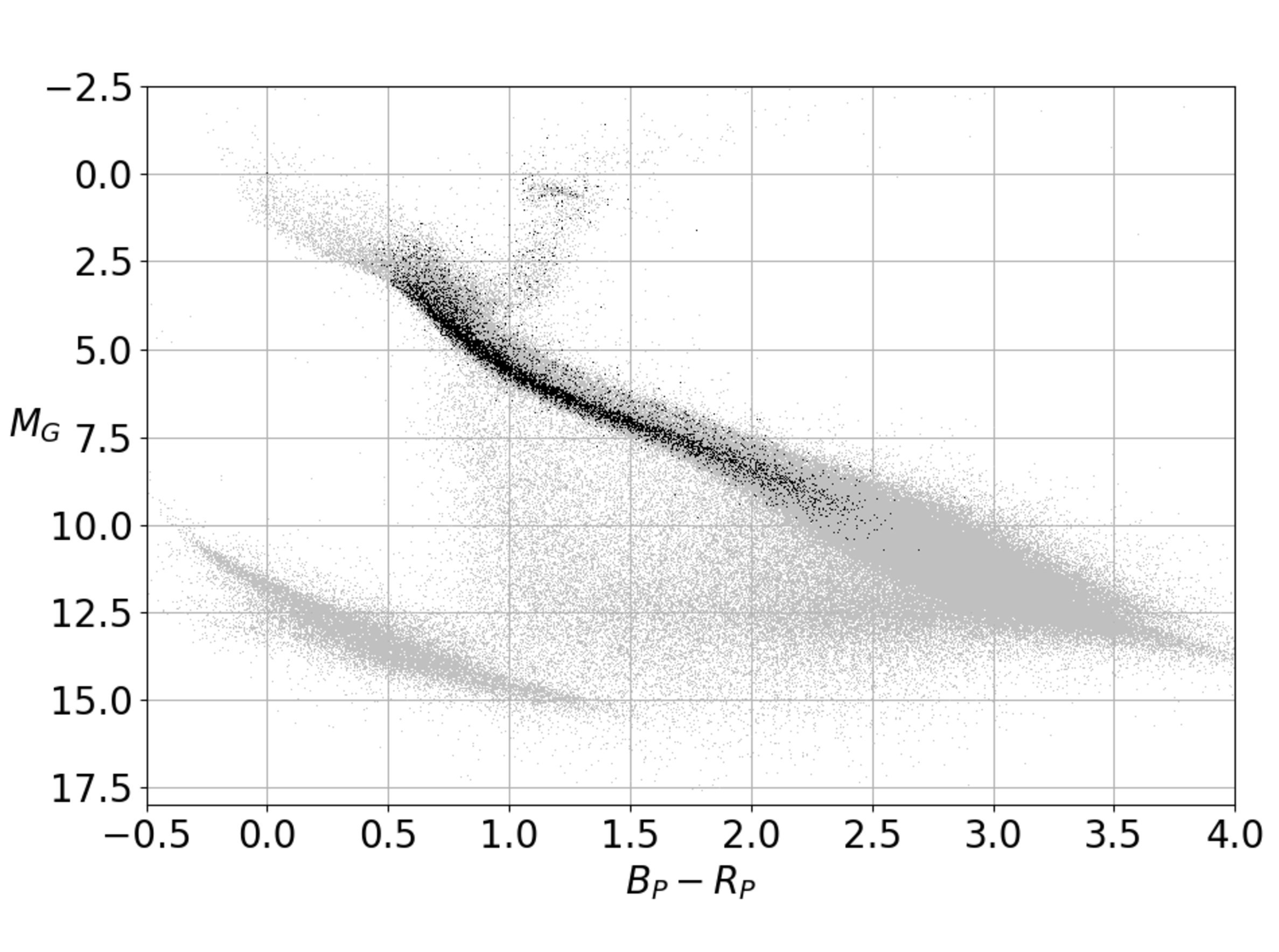}
     \hspace*{5pt}
   \includegraphics[height=6.0cm,width=\columnwidth]{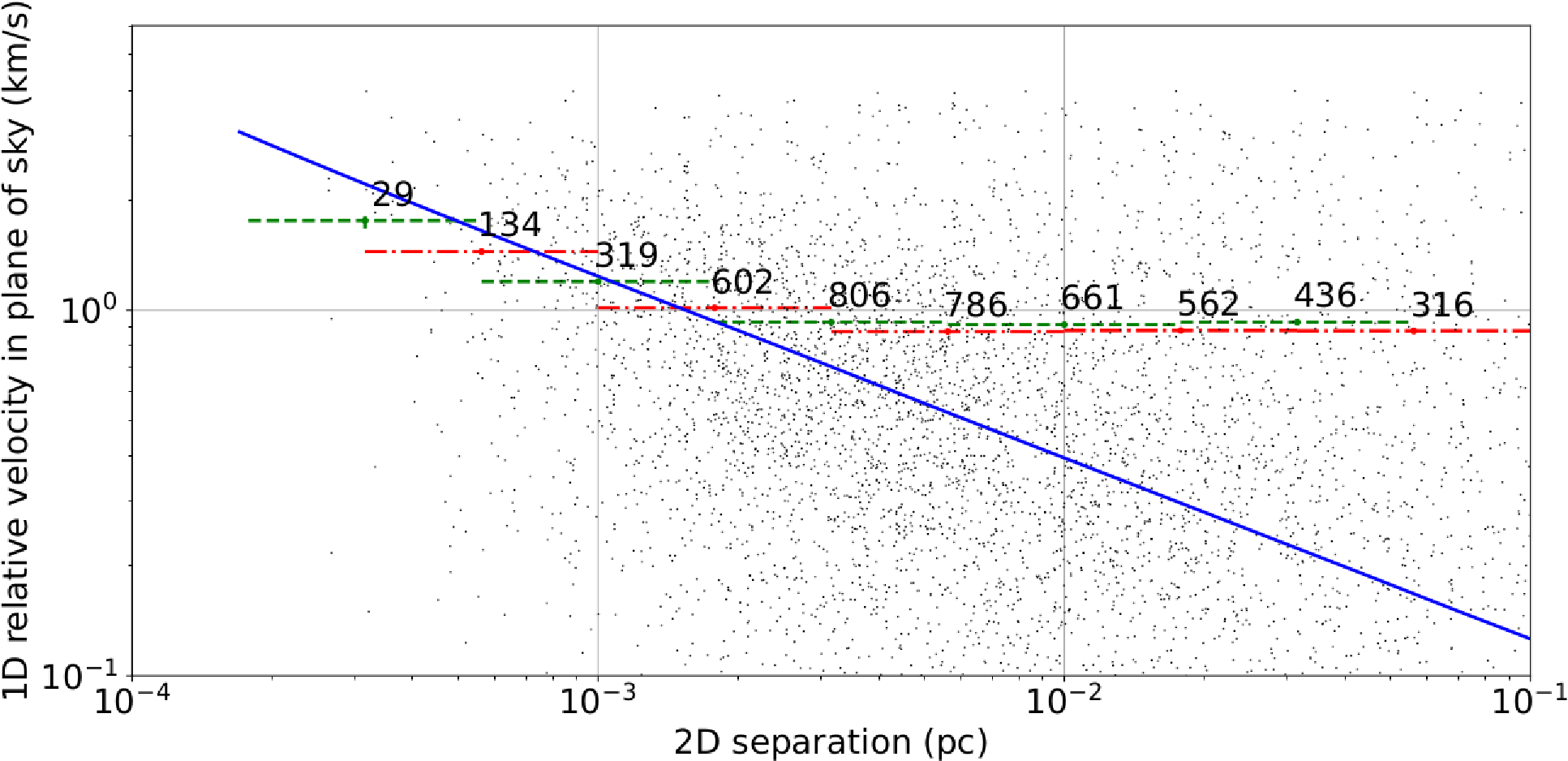}
   \caption{{  Left(a)}:Colour magnitude diagram for all the 5,736 stars of the 2,868 binary pairs shown in figure 3, also having relative
     radial velocities between both components < 4 km/s and signal to noise ratio in proper motions for both R.A. and Dec. for both
     components > 40, black dots, with the light grey dots showing the 169,540 stars of figure 2. The exclusion of white dwarfs,
     white dwarf/main sequence blends and the noisier low brightness main sequence is clear.
     {  Right(b)}: Root mean square binned distribution of 1D relative velocities between the members of the 2,474 binary pairs
     shown in the
     left panel of this figure, which also satisfy having relative velocities in both R.A. and Dec. < 4 km/s, and a minimum signal
     to noise in both of these quantities > 0.3, as a function of {  2D} projected separations between the members of each binary pair,
     for R.A. and Dec. measurements, dashed and dotted lines, respectively, {  with average relative errors for these two quantities
     being 10.77 and 11.85, respectively}. Each binary pair also appears as a pair of points at a fixed separation, for its
     corresponding relative velocities in R.A. and Dec. The solid line gives the Newtonian predictions of Jiang \& Tremaine (2010)
     for this same quantity.}   
 \end{figure*}

Our initial GAIA search returns all stars within 200 pc of the sun having precise parallaxes (signal-to-noise > 20). The query then scans
a projected circle of 0.5 pc about each of these stars for potential companions also
required to have fairly precise parallax measurements (signal-to-noise > 5). Each potential binary pair is then rejected if the parallaxes
of the two stars result in a distance difference along the line of sight for the two components larger than twice the projected
separation, $s$, between them, at more than $3\sigma$, i.e., a potential binary pair must satisfy $\Delta d-2s <3 \sigma_{\Delta d}$.
In going to large distances, a growing fraction of close companions will not be
detected due to the fixed GAIA resolution, our sample is not complete in any volume-limited sense, the requirement is not
than we should not miss any valid candidates, but that we should not include invalid ones. This initial search returns close to one
million potential binary pairs shown on a sky plot in figure 1.

It is clear that a number of well known
local groupings and associations have remained, where wide binaries will not satisfy any strict isolation criterion. Also, crowded regions
following the disk of the Milky Way are evident. Next, we demand that if any particular star appears as either the primary or the secondary
of more than one potential binary pair, all such potential binary pairs are removed. This results in an isolation sphere of at least
0.5 pc about all of the potential binary pairs passing this criterion, and allows to minimise the effects of possible
dynamic perturbations due to other nearby single stars forming part of the average stellar field population. In limiting
ourselves to kinematic observations of binaries with projected separations below 0.1 pc, we select stars having no
other close neighbours with parallax measurements within at least 5 times the binary separation; this isolation
factor grows linearly in going to closer binaries.

After removing all binaries where either of the two stars forms
part of more than one such pair we are left with 169,540 binary pair candidates, shown in a full sky plot in figure 2. {  We can
see} that the strict cuts applied have removed all evident local groupings, as well as excised the crowded zones along most of the
Milky Way disk region.

Before examining the relative velocity distribution of the binary candidates, we will use the relative radial velocities of the two
components to minimise the presence of unbound pairs by requiring the difference in radial velocities between both components of each
binary to be below 4 km/s. Thus, we remove from consideration all binary pairs where either component is lacking a radial velocity
determination. Also, as pointed out in El-Badry (2019), ignoring detailed spherical geometry correction effects will lead to
spuriously inflated (and raising with $s$) relative velocities for near and wide binaries. Although this correction is only a minor contribution,
keeping only stars with radial velocity observations allows the inclusion of full spherical geometry corrections (e.g. Smart 1968) to adequately
estimate the true relative velocity of the binaries in our sample. Excluding grouped binaries and those where either component is missing a
radial velocity measurement effectively remove over 99\% of the potential binary candidates. Figure 3 shows the full sky plot of the 5,844 binary
pairs remaining, a very significant reduction from the candidate sample of figure 2.

We do introduce
two modifications to the El-Badry \& Rix (2018) query, the first a slight increase in the radius of the projected circle
defining the initial search, from the $5 \times 10 ^{4}$  AU $= 0.25$ pc used by those authors to the 0.5 pc considered here, so as to
allow us to explore a somewhat larger range of binary separations. Finally, El-Badry \& Rix (2018) introduce a cut in relative velocities
on the plane of the sky to exclude any potential binary candidates which are inconsistent with Newtonian dynamics for a $5 M_{\odot}$
total mass binary to $3\sigma$. As our aim is to explore precisely the presence of wide binaries showing relative velocities
above Newtonian values, we remove this last cut.

Then, we apply strict signal-to-noise {  ($>40$)} quality cuts to the proper motion measurements in both R.A. and Dec. for both of the components
of each potential binary, in looking for a small sample where all stellar properties used are highly accurate. Only potential binary
pairs were both components pass the above data quality cuts in both of their proper motion measurements are retained. In order to minimise
the presence of chance associations resulting from random pairs of stars drawn from the local field population, we introduce
{  the cut mentioned previously} requiring that the radial velocity difference between the two components of all potential binary
pairs must be of less than 4 km/s, these last cuts leave us with 2,868 binary pairs. This relative velocity limit takes the place of
the Newtonian consistency criterion in the original query, and ensures that our $\Delta V$ inferences will be minimally affected by
random fleeting encounters of field stars, which would have velocities each drawn from a Gaussian distribution having a velocity
$\sigma$ of around 40 km/s, and hence a pairwise velocity difference having also a Gaussian distribution with $\sigma=\sqrt{2} \times$
40 km/s $\approx 60$ km/s.

 \begin{figure*}
     \includegraphics[height=6.0cm,width=\columnwidth]{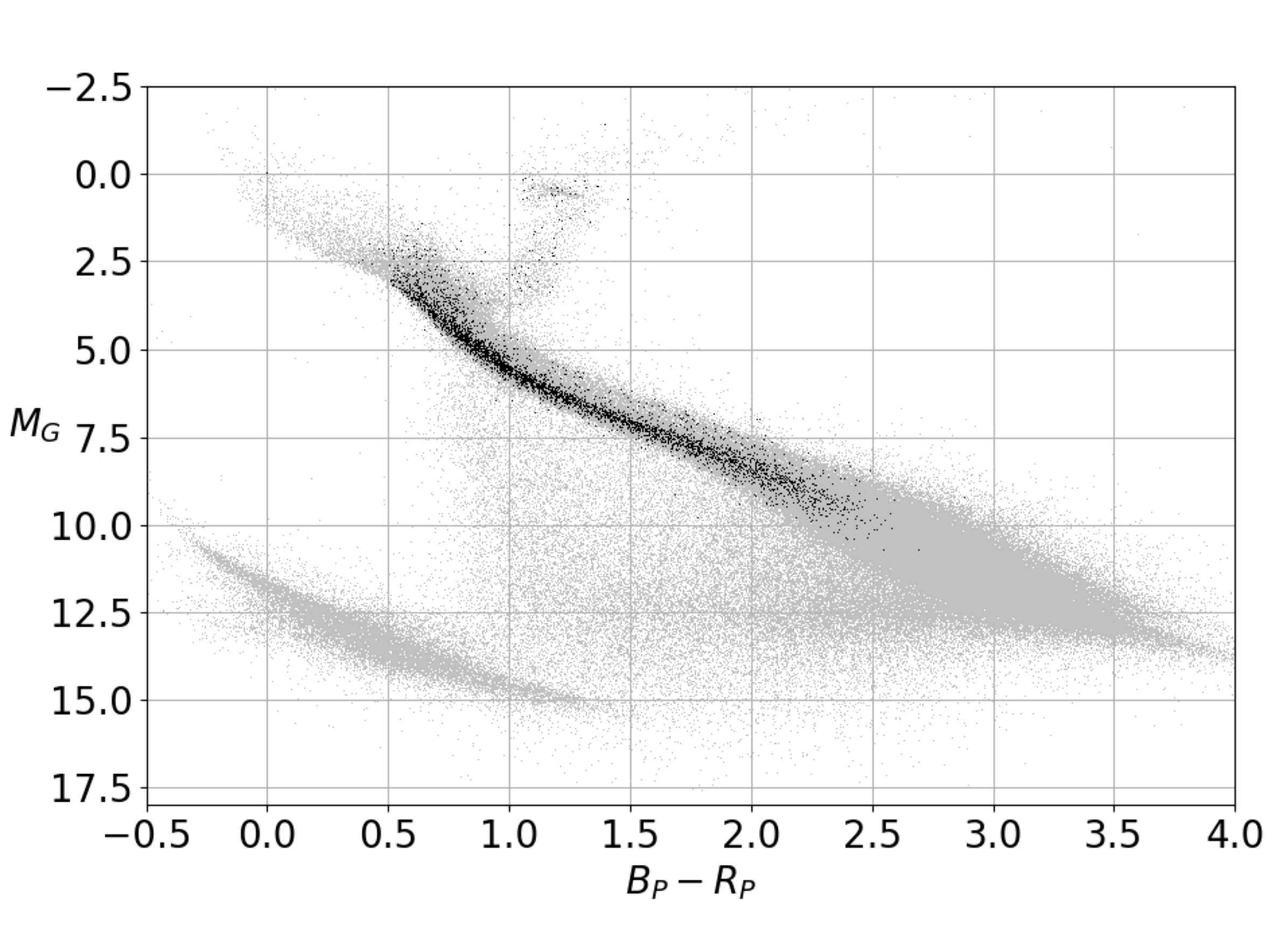}
     \hspace*{5pt}
    \includegraphics[height=6.0cm,width=\columnwidth]{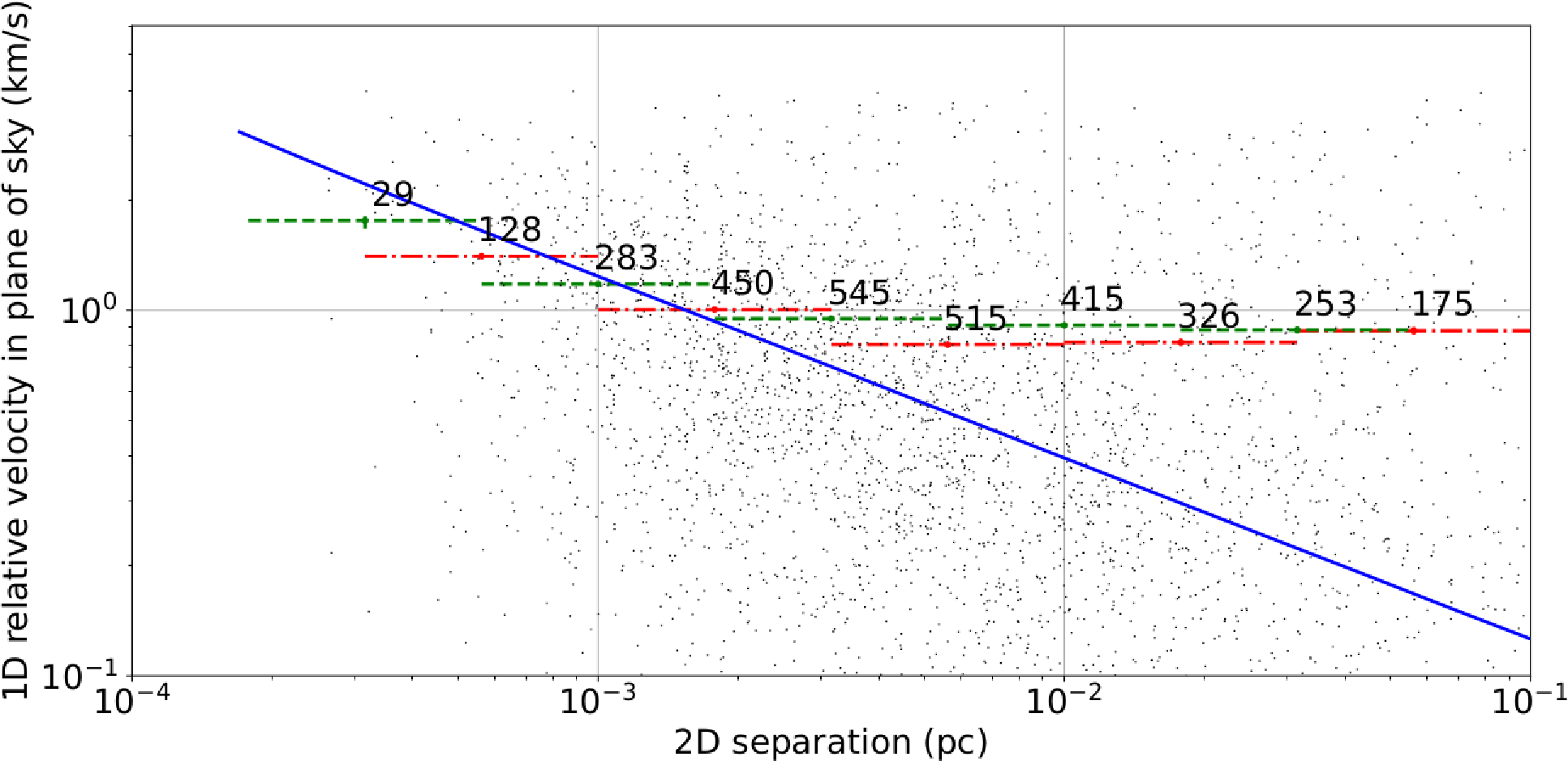}
    \caption{{  Left(a)}:The figure is analogous to the left panel of figure 4, but contains only {  as black points} the 3,700
      stars from the 1,850 binary pairs remaining after the further
      inclusion of a distance cut of 130 pc from the sun, resulting in a significant reduction in the numbers of photometric binaries.
      {  Right(b)}:The figure is analogous to the right panel of figure 4, but this time contains only 1642 binary pairs, after the
      distance cut described above. {  Average relative errors for right ascension and declination velocity differences are of 
      13.17 and 15.01, respectively.}}
 \end{figure*}

 Figure 4a gives the colour-magnitude diagram for this initial sample, dark black points for the 5,736 stars from 2,868 binaries remaining,
 with the 169,540 stars from the de-grouped
binary pairs, but having no radial velocity measurements shown as faint grey points. As in all subsequent colour-magnitude diagrams,
each binary pair contributes two points, by construction, of the same shade. It is clear that the stars having no radial
velocity measurements, are systematically dimmer than the brighter ones where the GAIA eDR3 catalogue returns also radial velocities.
The former include the dimmer solar neighbourhood main sequence, white dwarfs to the lower left hand region of the plot, and even a
region of white dwarf/main sequence unresolved blended binaries as a diffuse cloud between the main sequence and the white dwarf
population.

The main sequence for the selected stars is obvious, as is a band of {  narrow} photometric binaries just above the clear main sequence
composed primarily, but not exclusively, of grey points. Indeed, most photometric binaries do not have radial velocity measurements
in GAIA as the single star solution fails, providing an extra criterion for excluding such blended cases. As we infer relative velocities
on the plane of the sky through proper motion observations {  over the duration of the GAIA eDR3 catalogue, rather than through instantaneous
Doppler shift inferences}, we are not overly sensitive to either accelerations or velocities induced by hidden tertiaries, only to resulting
displacements, which to a varying degree, will be averaged out over the orbit of the hidden tertiaries. {  Very tight hidden
tertiaries with orbital periods much shorter than the 34 month eDR3 GAIA mission, lead to high instantaneous velocities, but to very
low displacements, which hence imply negligible kinematic contamination through our proper motion relative velocity determinations.
Of course, for hidden tertiaries of larger separations, a contamination effect inflating the wide binary relative velocity inference
appears, the following section is dedicated to as careful a strategy to minimise hidden tertiaries in
our final sample as the current data allow.}

We end this section with a kinematic plot for the initial sample in figure 4b, a rms binned distribution of
1D relative velocities as a function of 2D projected binary separation, dashed bins for R.A. measurements, and dotted for Dec. ones.
Further removal of binaries with final relative velocity signal-to-noise values $<0.3$ or $\Delta V >4$ km/s on either R.A. or Dec.
leaves us with 2,474 binary pairs having mean signal-to-noise values for R.A. and Dec. proper motions of 3,314 and 3,285, and
mean signal to noise for parallax measurements of 683. These last two cuts will be also implemented in all of the following kinematic
distributions. The mean distance to the Sun for these stars is of 107 pc, and the mean RUWE parameter 1.34. The last low signal-to-noise
lower limit on inferred velocities mentioned only excludes a few noisy outliers, the mean signal-to-noise values for this sample are
of 10.77 and 11.85 for R.A. and Dec. velocity inferences, respectively. 

As we have ended the sample considered at separations of 0.1 pc, and given the exclusion sphere introduced in the sample selection of
0.5 pc, our binaries included having the largest separations are isolated from any other GAIA eDR3 source having parallax measurements
out to 5 times the binary separation. This isolation factor grows to 50 times the binary separation for separations of 0.01 pc, and
continues growing linearly towards the smaller separations considered.

The solid line gives the results
of Jiang \& Tremaine (2010) for the rms 1D relative velocities for a simulated population of 50,000 binary pairs composed each of
two one solar mass stars, and for a random distribution of line of sight projections and a reasonable distribution of ellipticities,
after 10 Gyr of evolution in the solar neighbourhood under the influence of random encounters with field stars and the influence
of the galactic tidal fields, assuming Newtonian gravity. This line closely follows a Keplerian scaling of
$\Delta V_{RMS} \propto s^{-1/2}$ out to the tidal radius of the problem, at close to 0.7 pc, lying beyond the region being explored here.
Indeed, we end the figure at a separation of 0.1 pc, which implies that for all the binaries included, both components are at least
5 times closer than any other GAIA eDR3 source from the initial selection. Dim sources having poor parallax measurements can exist
within the 0.5 pc projected circle about each of the binary components, but mostly, given our small 200 pc distance cutoff,
will be background objects.

We see in figure 4b kinematics consistent with the Newtonian prediction of Jiang \& Tremaine (2010) for the lower separation bins,
and a transition to a constant relative velocity distribution for the $s>1.5 \times 10^{-3}$ region. The faint points in this
figure give each individual measurement, as in all subsequent kinematic plots, a small fraction of very small velocity individual
points are not displayed as they fall below the range selected to allow a full presentation of the results. The points shown can not
be individually compared to the Newtonian predictions of Jiang \& Tremaine (2010) shown, as this prediction refers to the rms values
of the resulting distribution of individual points, which is what the numbered binned values given present.
\footnote[1]{  The rms value for
the relative velocities for a large sample of wide binaries modelled under Newtonian gravity and having a distribution of ellipticities
and orbital projections with respect to an observer from Jiang \& Tremaine (2010), given by the solid line in figure 4b,
can only be compared to the rms values of distributions of relative velocities for wide binaries, not to individual values for any
particular wide binary. Thus, we give the individual points for the wide binaries obtained to show the full distribution of values
we obtain, not as a comparison to the solid line, which can only be meaningfully compared to the binned rms values given. Notice also
that the rms value of a distribution is skewed towards large values with respect to the mean, such that if one has a distribution of
points appearing equally distributed above and below a prediction for an rms value, the distribution being considered will necessarily
have an rms value larger than the prediction in question.}

 \begin{figure*}
     \includegraphics[height=6.0cm,width=\columnwidth]{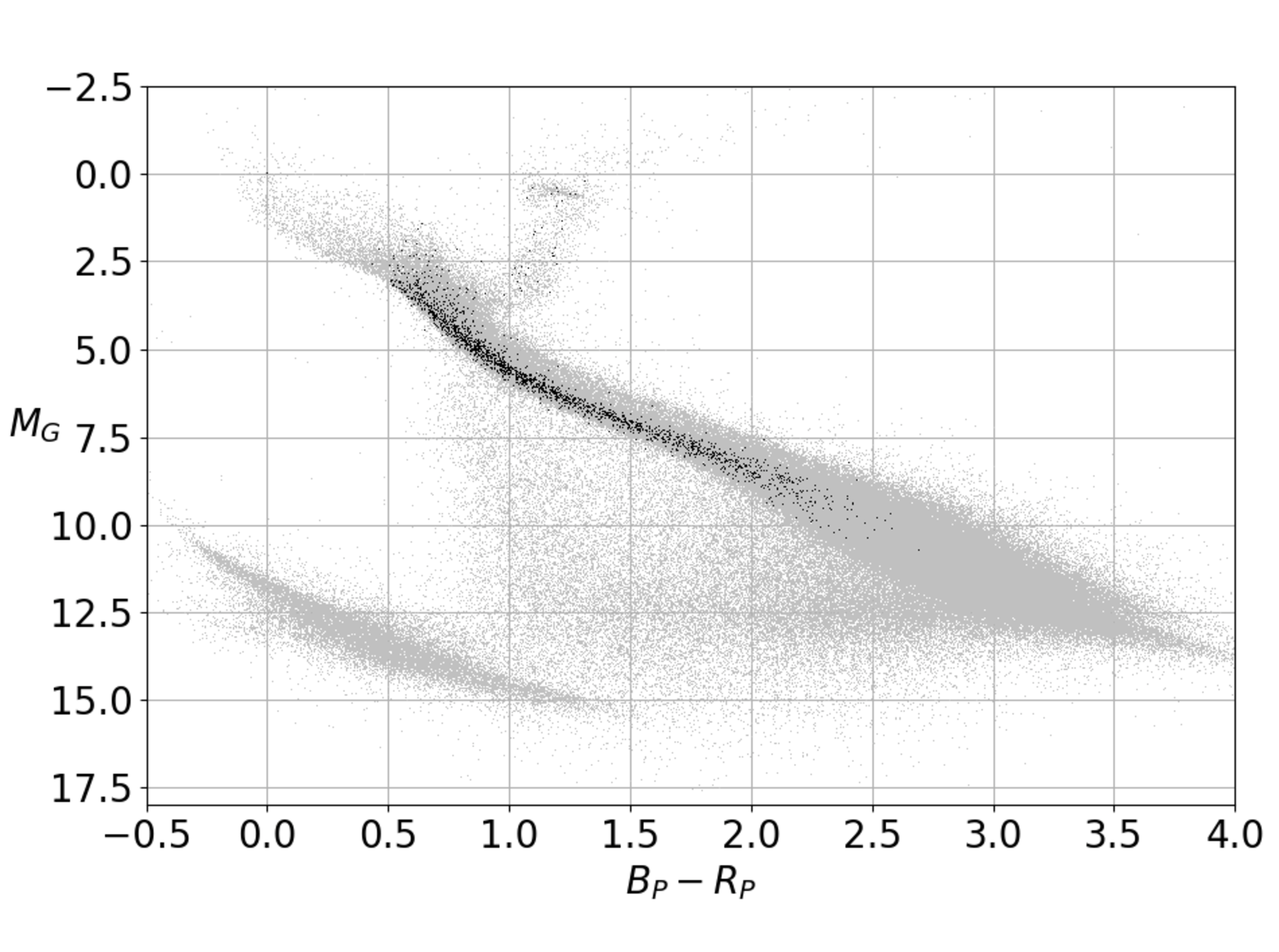}
     \hspace*{5pt}
     \includegraphics[height=6.0cm,width=\columnwidth]{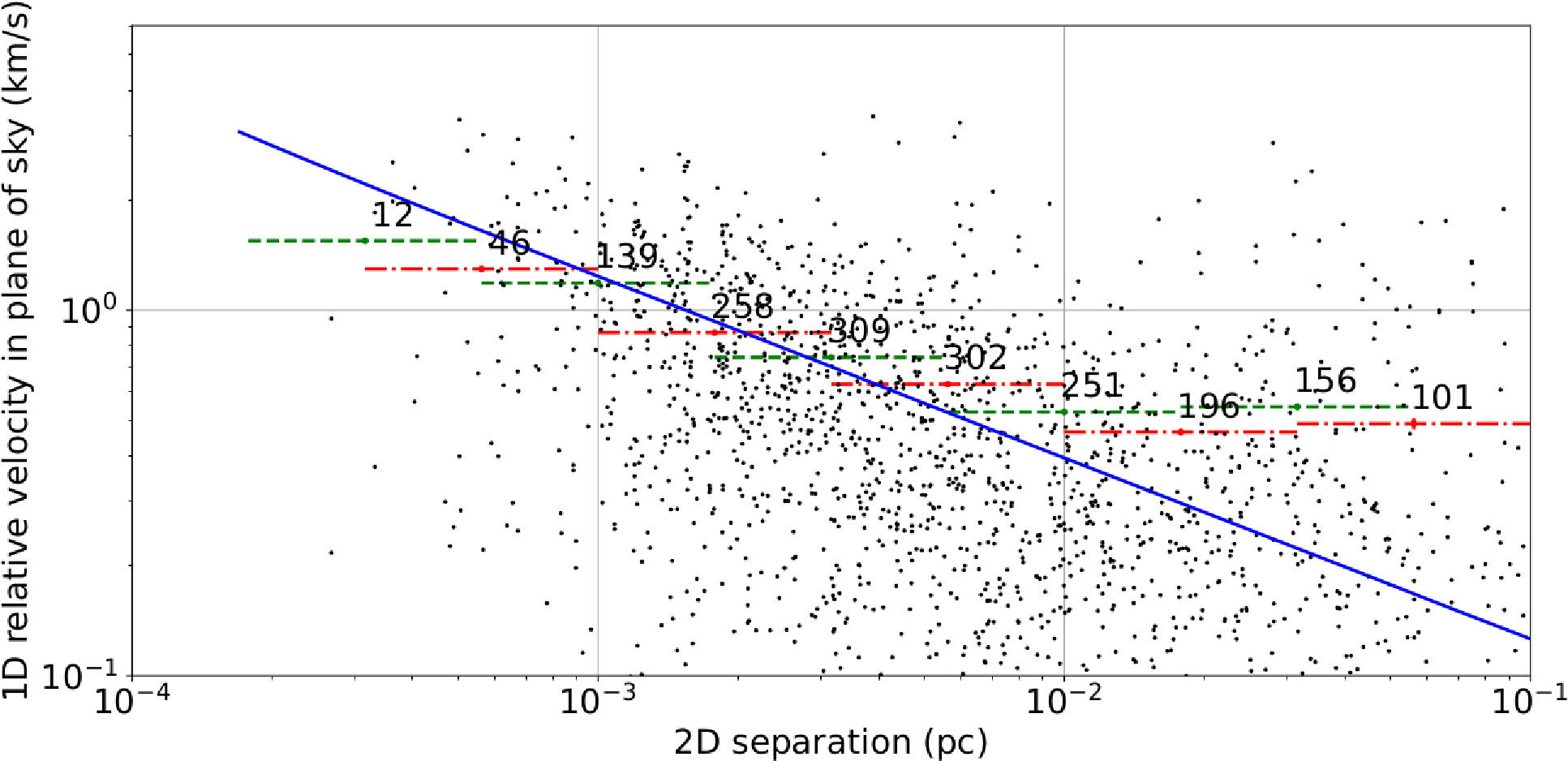}
     \caption{{  Left(a)}:The figure is analogous to the left panel of figure 5, but contains only as black points 2,016 stars from the
       1,030 binary pairs after the further inclusion of a RUWE < 1.2 cut, resulting in an almost complete exclusion of photometric
       binaries and a significant reduction in the number of post turn-off stars.
      {  Right(b)}:The figure is analogous to the right panel of figure 5, but this time contains only 929 binary pairs, after the
      data quality cut described. {  Average relative errors for right ascension and declination velocity differences are of 
      14.39 and 15.88, respectively}.}
 \end{figure*}

A more careful exploration of the physics behind these trends can be attempted by considering an estimate of the total mass of {  each
binary} in this initial sample, and a study of the mass-velocity scalings present. We use the mass estimate in Pittordis \& Sutherland
(2019) of:

\begin{equation}
\left( \frac{M}{M_{\odot}} \right) = 10^{0.0725(4.76-M_{G})},
\end{equation}

\noindent where the G band GAIA absolute magnitude is shown to provide a good stellar mass determination though the above equation,
for main sequence stars. Power law fits to binned distributions of average $\Delta V$ as a function of binary mass for the data shown
in figure 4b yield logarithmic slopes of $0.33 \pm 0.12$ and $0.32 \pm 0.07$ for the $s<1.5 \times 10^{-3}$ and $s>1.5 \times 10^{-3}$
regions, respectively. The lack of a clear $\Delta V \propto M_{b}^{0.5}$ scaling in the $s<1.5 \times 10^{-3}$ region probably indicates
the presence of significant kinematic contamination in our initial sample, as evident for example in the presence of a few spectroscopic
binaries in the black points of figure 4a. Also, the consistency of the two mass-velocity scalings mentioned above suggests a common
origin for the relative velocities shown, again probably dominated by kinematic contaminants. Further cleaning of the sample to minimise
the presence of the above and other sources of kinematic contamination is described in the following section.

\section{Clearing kinematic contaminants}

In this section we describe a sequence of three further data quality and kinematic contaminant exclusion criteria, which will
be applied sequentially, such that at any point, all previous criteria are also present.
The first step towards eliminating kinematic contaminants consists of introducing a distance cutoff to restrict our sample to
a smaller, higher quality set where the prevalence of undetected stellar companions diminish, as all apparent magnitude dependent
systematics are reduced. Of course, any restriction in the maximum distance considered implies a trade-off in terms of a fast drop
in the number of remaining binary pairs. We find that the most restrictive distance cutoff which still yields a workable number of
binary pairs is 130 pc. This further restriction leaves us with 1,850 binary pairs, the 3,700 stars shown in the colour-magnitude
diagram of figure 5a, with the background light grey points being the same as the ones shown in figure 4a.

The corresponding kinematic plot is presented in figure 5b, where after exclusion of low velocity signal to noise pairs and imposing
a maximum relative velocity on the plane of the sky of 4 km/s, as described in the case of figure 4b, we are left with 1,642 binary
pairs. The mean distance, mean signal to noise in R.A. and Dec and mean RUWE values of this sample are of 79.6 pc, 4,091, 4,047 and
1.37, respectively. We see the reduction in the distance resulting in more accurate astrometric observations.

Comparing the colour-magnitude diagrams of figure 5a to figure 4a, we see that the reduction in the distance significantly reduced the
number of photometric binaries in the sample in the distinct band just above the main sequence, which also appears more clearly defined.
Brighter stars in the turn-off region are also diminished, as the total numbers dropped. In comparing figures 4b and 5b we see that the
distance reduction and corresponding kinematic cleaning of the sample resulted in a slight drop in the amplitude of the constant
$\Delta V_{RMS}$ region to the right of the plot, but no significant qualitative changes. Indeed, the mass-velocity scalings for both
the region following the Newtonian predictions of Jiang \& Tremaine (2010), and the one to the right of this presenting the slight
drop in amplitude described above, show again power law fits with logarithmic slopes consistent with those of figure 4b, and not
corresponding to any evident physical scalings.

 \begin{figure*}
    \includegraphics[height=6.0cm,width=\columnwidth]{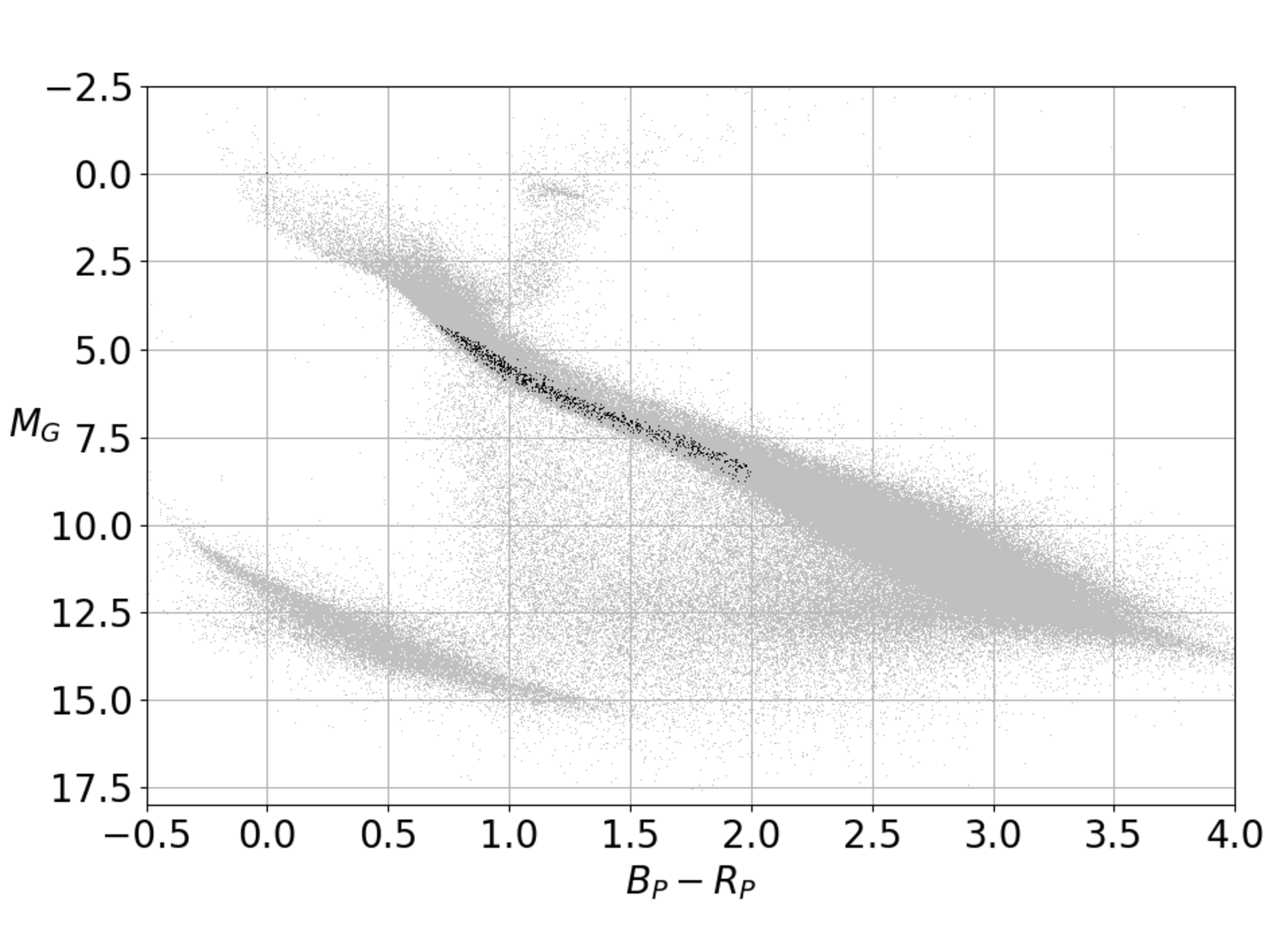}
     \hspace*{5pt}
    \includegraphics[height=6.0cm,width=\columnwidth]{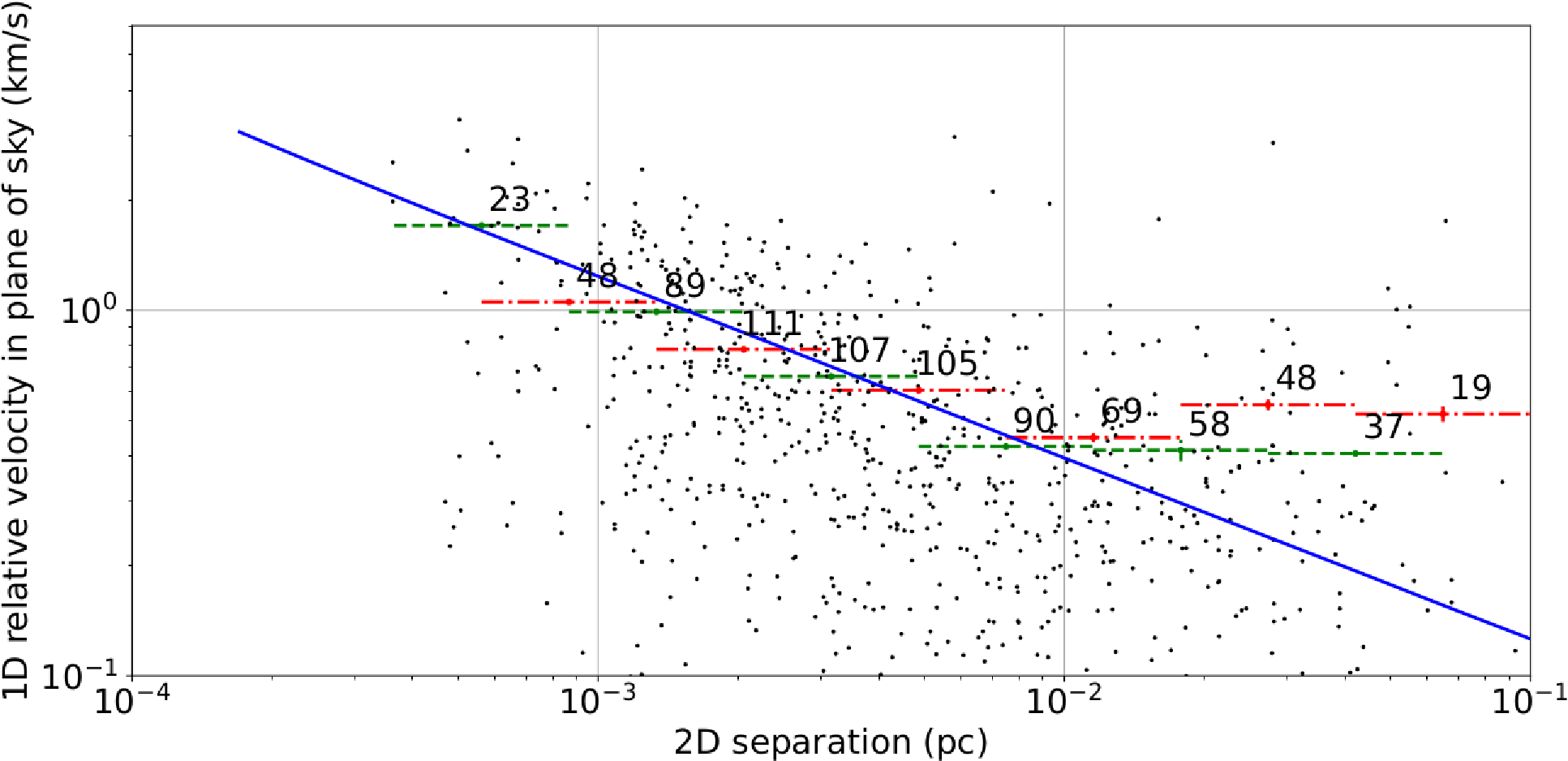}
    \caption{{  Left(a)}:The figure is analogous to the left panel of figure 6, but contains only as black points 888 stars from the
      444 binary pairs after the further inclusion of a strict colour-magnitude selection, as shown in this figure, restricting the sample
      to the cleanest region identified by Belokurov et al. (2020) in terms of near absence of any kinematic contaminants. Any binary
      pair where either component falls outside of the selected region is removed.
      {  Right(b)}:The figure is analogous to the right panel of figure 5, but this time contains only 423 binary pairs, after the
      data quality cut described. {  Average relative errors for right ascension and declination velocity differences are of 
      14.88 and 18.62, respectively}.}
 \end{figure*}

The next quality cut introduced is to directly limit the maximum of the allowed RUWE parameters for the stars considered. This parameter
gives a measure of the goodness  of fit for the GAIA single star solution performed on each of the sources modelled, and is known to shift
to larger values in cases where unresolved stellar companions are present, e.g. Belokurov et al. (2020). We introduce an upper cut
of RUWE < 1.2 such that if either the primary or the secondary of a given binary pair fails this test, the binary pair is removed from the
catalogue. Comparing to the mean RUWE parameters of the samples displayed in figures 4 and 5, of 1.34 and 1.37, it is clear that the upper
limit introduced at this stage presents a fairly large cut.

In fact, the black points in the colour magnitude diagram of figure 6a represent only the 2,016 stars from 1,008 binary pairs showing
an extremely well defined main sequence region with practically no evident photometric binaries above it. Also, the turn-off region is
now almost empty, as is also the case for the main sequence to the right of a colour of about 2.25, where larger errors, and hence
higher probabilities of contamination from unresolved stellar companions, are evident from the underlying light grey distribution.

 \begin{figure*}
    \includegraphics[height=6.0cm,width=\columnwidth]{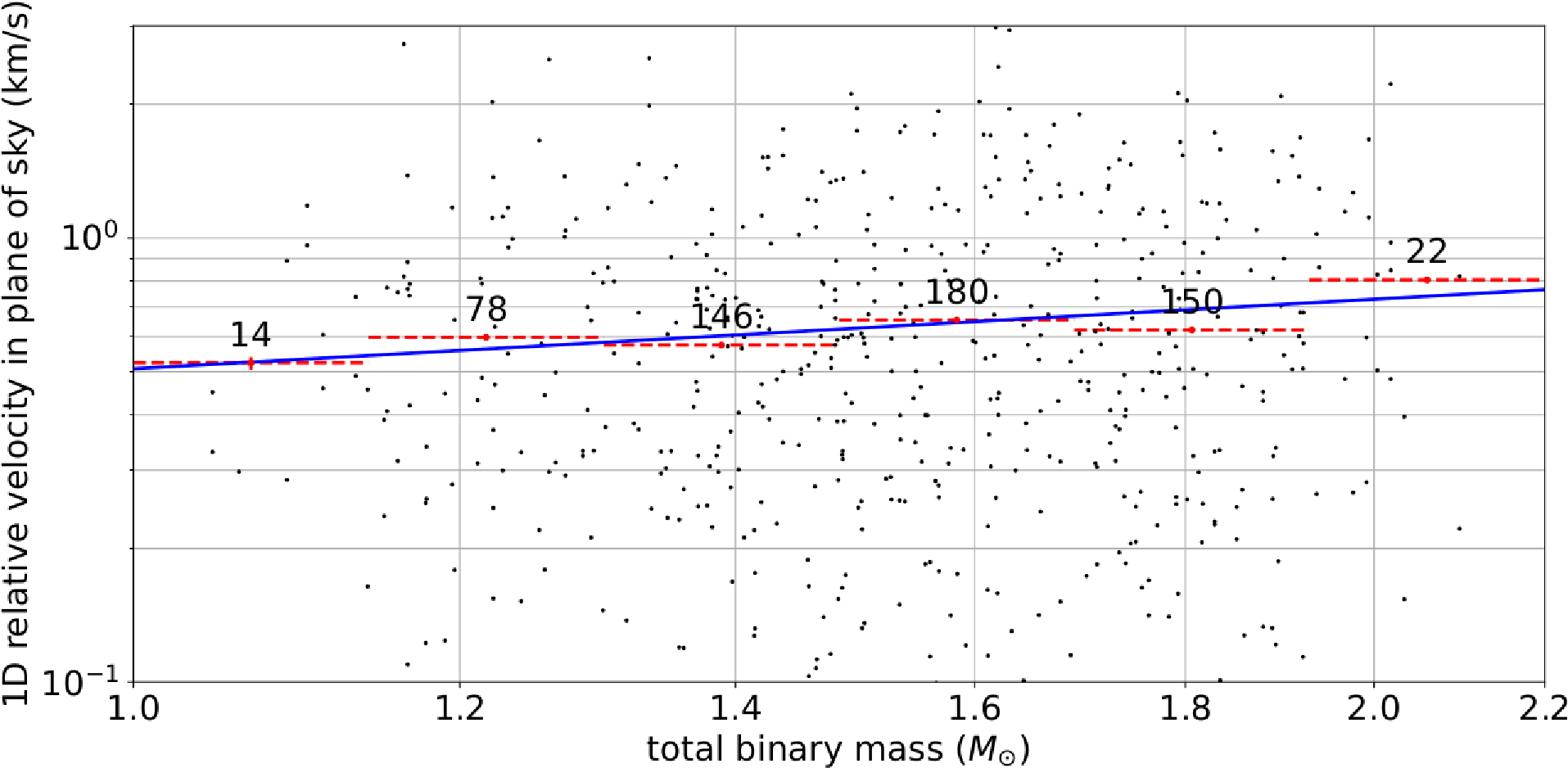}
     \hspace*{5pt}
    \includegraphics[height=6.0cm,width=\columnwidth]{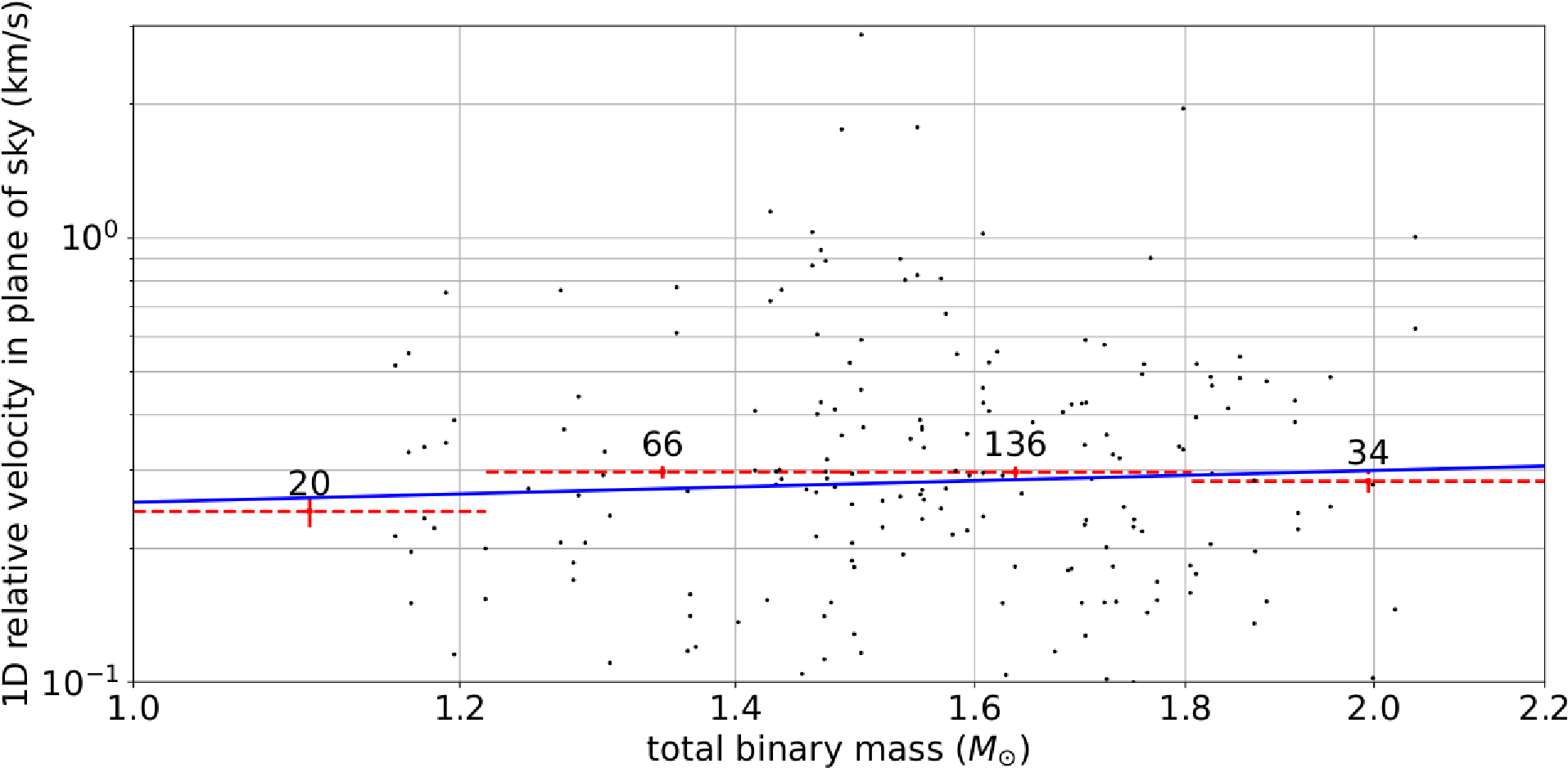}
    \caption{{  Left(a)}: This panel shows the binned average 1D relative velocity vs. binary mass scaling for the binaries with
      separations <0.009 pc in the right panel of figure 7, were each binary pair contributes two points, one for R.A. and one for Dec.
      observations. {  Average relative errors for right ascension and declination velocity differences are of 
      18.36 and 23.56, respectively}. A power law fit to this scaling yields $\overline{\Delta V} \propto M_{b}^{0.52 \pm 0.14}$.
      {  Right(b)}: This panel is analogous to the left one of this figure, but for separations >0.009 pc. {  Average relative errors
      for right ascension and declination velocity differences are of 6.85 and 7.24, respectively}. A power law fit to this scaling yields
      $\overline{\Delta V} \propto M_{b}^{0.24 \pm 0.21}$.}
 \end{figure*}

Figure 6b gives the kinematic plot for this sample, which contains only 929 pairs of wide binaries having very high quality GAIA
fits for both components, indeed, the mean RUWE values this time are of only 0.99. As a reflection of having a much cleaner sample
than we started with, the mean signal-to-noise values for proper motion observations in R.A. and Dec., and parallax are now
4,123, 4,073 and 911. This time we see a more pronounced change with respect to the previous figures 4b and 5b, in that the
region over which our results trace carefully the Newtonian predictions of Jiang \& Tremaine (2010) now extends to larger separations,
with results being consistent with it out to close to $10^{-2}$ pc. Concurrently, the amplitude of the constant $\Delta V_{RMS}$ region
has again dropped and now appears at about 0.5 km/s. Although the total numbers have dropped considerably from the 2,474 binaries of
figure 4b, due to the removal of binary candidates where the probabilities of kinematic contamination were much higher than for the
remaining ones, the very small dynamical range in the total masses, of only about a factor of 2.2, still does not allow a clear detection
of any physically significant mass velocity trend, over any of the regions in this diagram.

 We now turn to the final pruning of our catalogue where we select directly on the colour-magnitude diagram for regions containing minimal
 presence of unresolved stellar companions, after all the cuts described previously have been sequentially applied. Belokurov et al. (2020)
 present an extremely detailed analysis of binarity, variability and kinematic contamination on wide binary relative velocities as
 a function of RUWE parameter and location in the colour-magnitude diagram of the GAIA catalogue. Making use of their results, we define
 a narrow region of the upper main sequence as the final inclusion criteria. The region selected, morphologically through $M_{G}$
 and $B_{p}-R_{p}$ ranges, in addition to our previous RUWE filter, limits the presence of any kinematic contaminants through unresolved
 stellar companions to less than $5\%$, according to the results of Belokurov et al. (2020), e.g. their figure 10 {\it right} for the
 magnitude range we use here. Indeed, even the presence of hot or outer Jupiters can be constrained to fall rapidly for stars with RUWE
 indices below 1.0. As with all our previous cuts, all binaries for which either the primary or the secondary fail the colour-magnitude
 selection criteria, are removed.

 Our inclusion region in the colour-magnitude plot is shown in figure 7a, and comprises only a very well defined section of the
 upper main sequence. This region is defined as all stars within a $\pm$0.4 vertical magnitude interval of the line joining points
 (0.7, 4.7) and (2.0, 8.7), where the numbers give magnitude, colour coordinates. This final selection leaves only 444 binary pairs,
 which after the kinematic signal to noise and upper relative velocity criteria of 4 km/s, results in the 423 binary pairs appearing
 in figure 7b. For this final high-quality sample the average signal-to-noise values for the GAIA input parameters used are now
 of 4,202, 4,064 and 950 for R.A. and Dec. proper motions and parallaxes, respectively. The final average relative velocity
 signal-to-noise values for this sample are {  about 50$\%$ higher than} those of the initial sample of figure 4, 14.88 and
 18.62 for R.A. and Dec., respectively.

 This final kinematic plot is largely consistent with the previous one of figure 6b, with the amplitude of the region showing no
 dependence of the $\Delta V_{RMS}$ values on separation still at 0.5 km/s. Also evident is the appearance of random fluctuations
 indicative of shot noise beginning to become relevant as the numbers of binary pairs considered have continued to fall. The region
 of consistency with Newtonian predictions now extends to about $10^{-2}$ pc.

 Finally, we present in figure 8a the mass-velocity scaling for the 295 binary pairs having separations below 0.009 pc in figure 7b,
 and the 128 such pairs with separations larger than 0.009 pc also from figure 7b, in figure 8b. It is interesting that despite the very
 narrow dynamical range allowed by intrinsic stellar physics and the low numbers remaining after the very strict series of cuts applied,
 figure 8a shows a scaling of mean 1D velocity differences for the selected binaries of $\overline{\Delta V} \propto M_{b}^{0.52 \pm 0.14}$
 (with a correlation coefficient of 0.88, {  much higher than that of any previous mass-velocity scalings}), perfectly consistent with
 Newtonian expectations, which indeed are clearly met as apparent in
 the $\Delta V_{RMS} \propto s^{-1/2}$ scaling shown in figure 7b for this region. This last result validates the procedure undertaken aiming
 at identifying an extremely high quality and high purity catalogue of isolated solar neighbourhood wide binaries, albeit including only
 small numbers of such stellar pairs.

 Our final plot shows the mass velocity scaling for the 128 wide binary pairs having $s>0.009$ in figure 7b. The average binary mass
 for this last plot is of $1.6 M_{\odot}$, as it is also the case for figure 8a. The scaling in this final plot is of  $\overline{\Delta V}
 \propto M_{b}^{0.24 \pm 0.21}$, with a correlation coefficient for these small numbers of stars of 0.63. As with the previous plot, we have
 divided the full mass range into as many bins as permitted by the constraint of having at least 10 data points in each bin. {  The GAIA
 eDR3 identifiers and relevant parameters used for these 128 wide binary pairs appear in the appendix.}

 We caution that the confidence intervals quoted for the mass velocity
 scalings in all cases are lower limits to this quantities, as they are only formal statistical ranges which ignore a series of
 statistic and systematic errors still present. One of the above is the variance inherent to the stellar mass estimates of equation (1),
 which are hard to estimate. In the similar magnitude, colour mass estimate through isochrone comparisons of El-Badry \& Rix (2018), a
 confidence interval of $ \pm 0.1 M_{\odot}$ for their photometric mass estimates, which closely match those of equation 1 for the
 main sequence range we use, is given.

 Also, given the small numbers of binary pairs
 available in figure 8b, the shallow velocity mass scaling present and the narrow $M_{b}$ dynamical range accessible, the slope obtained
 is sensitive, to within about 1.5 times the quoted confidence interval, to details of the various data quality cuts and kinematic
 contamination cleaning procedures applied. Still, there is a clear indication for a lower slope for the velocity mass scaling in the
 $s>0.009$ pc region compared to the one found in the $s<0.009$ pc one, which is in fact consistent with the
 $V_{TF}=0.35 (M/M_{\odot})^{1/4}$ km/s galactic $a<a_{0}$ Tully-Fisher scaling. 

 {  To summarise, hidden tertiaries with bright companions are effectively photometric binaries, which are efficiently excluded
 through the HR diagram selection, as well as through their lack of accurate single stellar photometric and spectroscopic solutions,
 evident through large values of the RUWE parameter and lack of reported radial velocities, respectively. Hidden tertiaries
 with dim companions and separations lower than about 10 AU induce a wobble on the detected member of the large scale binary,
 which results in poor GAIA single stellar fits, and hence can be largely excluded through RUWE parameter upper limits.
 On this point, Belokurov et al. (2020) estimate less than 5\% hidden tertiaries contamination for internal separations below 10 AU,
 on the HR region of interest, for a RUWE upper limit of 1.4, for stars within 1 kpc of the sun. These
 numbers can be compared to the much stricter upper RUWE cut of 1.2 (and final average values of 0.99) and distance limit of 130 pc
 which we impose on our final sample. Hidden tertiaries with internal separations between about 10 and 100 AU will be hard to exclude,
 although their frequency will drop as the distance limit of our sample is reduced. Even assuming all of these appear in our sample,
 from empirical estimates of tertiary stellar systems (e.g. a total fraction of 40\% from Tokovinin et al. 2002, 2010) we can expect
 less than about a 10\% contamination in our final sample, for a equally populated logarithmic
 separation intervals, e.g. Clarke (2020).
 These systems would result in artificially enlarged relative velocities of $\lesssim$ 1 km/s
 on the small fraction of affected binaries. Finally, unbound flybys are restricted through the use of a relative radial velocity
 filter of $<4 km/s$, a relative velocity on the plane of the sky filter also of $<4 km/s$, and the careful de-grouping procedure
 which ensures only isolated binaries remain in our final sample. Given the mean average interstellar distance of
 1pc in the Solar neighbourhood, and the pairwise velocity dispersion of about 60 km/s for these stars, the presence of unbound
 flybys in our final sample of wide binaries with separations below 0.1 pc, is extremely unlikely.}
 
 Throughout this project we made use of \texttt{Bynary}, a suite of Python programs which easily downloads
 stellar and binary data from GAIA, filters and processes them via its colour-magnitude, Kinematic and Velocity/Mass modules to produce
 the graphs and statistics presented here, also allowing the user either to export data for other analytically programs and purposes or
 looking at the corresponding entries of any relevant binary pair in a range of possible surveys (e.g. DSS, 2MASS, AllWISE, GALEX,
 Fermi or IRIS). A full description and release of \texttt{Bynary} will be the subject of a future publication.

\section{Comparison to previous results}

We now present a comparison of our results to a series of recent similar studies, focusing on the degree to which these
are consistent with our current findings and the differences in sample selection criteria, data quality available and
assumptions made when interpreting the results obtained.

\subsection {Hernandez et al. (2019a)}

In Hernandez et al. (2019a) the data were much scarcer, as the authors started from a Hipparcos wide binary selection
sample by Shaya \& Olling (2011), leaving only 81 binaries in the final sample (in part due to the 1/3 missing Hipparcos sources in
DR1), with projected separations between 0.003 and 10 pc, allowing for only two bins below 0.1 pc. Thus, having one bin
at close to 0.04 pc showing a clear deviation from Newtonian expectations and one slightly below 0.01
where the large error bars made it ambiguous in terms of following or not the Newtonian prediction,
identified the region between 0.001 and 0.1 as crucial for detailed analysis. Figure 5 in Hernandez et al.
(2019a) does not allow more than an order of magnitude identification of the threshold beyond which
deviations from the Newtonian prediction appear, given the very limited number of bins and the large
error bars resulting from the small numbers of binaries available in that study.

Also, the large separations
sampled in that study become suspicious in terms of external perturbations, given the average interstellar
separations of close to 1 pc in the solar neighbourhood, the possibility of substantial kinematic contamination
due to interactions with surrounding stars becomes a worry beyond a few tenths of a pc. Extending the binary separation range
explored to close to 1 pc and beyond is problematic not only in terms of excluding external kinematic perturbations,
but also as binary identification becomes ambiguous, within the 1 pc typical interstellar separation of the
average distribution.

For the above, we decided here 
to concentrate on an extremely clean sample of isolated binaries where an exclusion sphere many times larger
than the binary separation could be defined about each, as detailed in the sample selection section, and
with significant resolution in the separation interval between 0.001 and 0.1 pc. These permit a clear tracing of
the Newtonian region, which serves as a control of the whole procedure in as much as an accurate agreement with
the Jiang and Tremaine (2010) predictions is recovered, and a more accurate identification of any divergence
occurring, within a regime highly free from contamination due to the average field star distribution. 

Given the small number of bins and much larger error bars in
Hernandez et al. (2019), our present results are consistent with those previous findings, to within the significantly
extended confidence intervals of that previous study.

\subsection {Clarke (2020)}

The highly relevant recent work of Clarke (2020) showed that the presence of blended tertiaries, cases where one or
both of the two components of an observed binary are in fact small scale binaries themselves containing an undetected
component, will result in kinematic contamination inflating the inferred relative velocities of the observed
large-scale binary. That study shows that a distribution of reasonable separations and masses of blended tertiary
components with a high hidden tertiary fraction of 0.5, can be invoked to fully account for the results of Hernandez
et al. (2019). It is interesting that such a distribution of undetected tertiaries should become apparent through a
variety of effects.

However, the results of Clarke (2020) depend critically on the details of the assumed distribution functions.
There, the separation distribution of hidden
tertiaries is assumed as uniformly distributed in log separation between 3 and 100 AU. This means
that a fraction of 0.343 of the hidden tertiaries in that study lie in the 3-10 AU interval.
This first inner separation interval is {  the one producing the greatest effect on the resulting
wide binary relative velocities inferences in the context of Clarke (2020), as it is the closest of
the hidden tertiaries which have the largest kinematic contamination effects on the large scale wide binary,
given the Newtonian $r^{-1/2}$ scaling for velocities with separation, and as all equal logarithmic intervals
were assumed there as equally populated. As Clarke (2020) acknowledges}, many of the hidden
tertiaries assumed in that study would imply anomalies in their location in the HR diagram, as well as astrometric
solutions of degraded quality, all of which allows to remove from consideration such affected binaries.
It is because of the result-defining effect blended tertiaries
could have on our analysis, that we have gone through an extremely thorough cleansing process,
guided by the Clarke (2020) results, and the Belokurov et al. (2020) criteria.

In that last paper,
through careful and extensive simulations within GAIA DR2 parameters and comparisons with observations
from GAIA DR2, it is concluded that close hidden tertiaries with separations below 10 AU can be
detected mostly through the RUWE flag, and also in combination with taking nearby samples with
high quality astrometric solutions, i.e. high signal to noise values in parallax and proper motions,
see also Penoyre et al. (2020). Taken together, the conclusion of the two above papers
is that with a RUWE filter of 1.4 and high signal to noise cuts for nearby samples, together with
HR selection criteria as we have applied, one can exclude hidden tertiaries to below a 5\% contamination
level, see e.g. the upper row in figure 9 of Belokurov et al. (2020) for the hidden tertiary fraction
for main sequence stars in the magnitude range we select. This excludes the 3-10 AU separation interval,
which is crucial for the Clarke (2020) result. Further, it is important to note that the Belokurov et al. (2020)
result is valid for the 22 month GAIA DR2 duration, the 1.55 {  times} larger 34 month eDR3 time interval, at comparable
data quality, translates into a $1.55^{2/3}=1.34$ larger separation exclusion of 13.4 AU. Thus, the original
$log(100/3)=1.523$ logarithmic interval assumed in Clarke (2020) should be reduced to a $log(100/13.4)=0.873$
interval, i.e. an interval smaller by a factor of 1.75, which crucially, is missing the inner 3-13.4 AU separation
range. Thus, in going from DR2 to eDR3, and taking all precautions derived from the results of
Belokurov et al. (2020) and Penoyre et al. (2020) we can be confident that close to half of the hidden
tertiaries assumed by Clarke (2020) to reproduce the non Newtonian high relative velocity distribution,
indeed, the most {  relevant} half, as the kinematic contamination effects of hidden tertiaries will scale
with separation as $s^{-1/2}$, are not valid as an assumption in our present data and experimental set up
to explain the results obtained at large separations.

Notice also that we have adopted a much more strict (in terms of the fraction of GAIA systems remaining)
RUWE cut of 1.2, compared to the fiducial value of 1.4 explored in Belokurov et al. (2020) as a reference
limit below which hidden tertiaries with separations below 10 AU can be largely excluded (for distances
of up to 1 kpc, much larger than the high quality limit of only 130 pc which we impose), and that the average
RUWE for our final sample is of a very stringent 0.99. It is clear that the hidden tertiary separation interval
allowed by our present set up 13.4-100 AU, would require significantly more than twice the hidden tertiary fraction
assumed by Clarke (2020) to explain the high relative velocity distribution found above Newtonian expectations.
As this assumed fraction in Clarke (2020) was actually of 0.5, already in the high end of empirical estimates
(e.g. of between 22\%-36\% per binary in Tokovinin et al. (2002), Tokovinin et al. (2010),
with no evidence of any separation dependence for this fraction), explaining our present results through the
Clarke (2020) argument would not be possible even if one assumes a limiting fraction of hidden tertiaries
of 1.

\subsection {Pittordis \& Sutherland (2019)}

Using Gaia DR2 Pittordis \& Sutherland (2019) examine the distribution of relative velocities between the components
of wide binaries, for four binary separation intervals between 0.025 and 0.1 pc. Thus, they sample the region where
we find a clear signal above Newtonian expectations. Just like in our present study, these authors find a clear
signal for relative velocities above Newtonian expectations, in an extended relative velocity distribution for all
their separation bins sampled, which they mention probably corresponds to the non-Newtonian signal reported in Hernandez
et al. (2019), and which corresponds to the kinematic signal modelled by Clarke (2020) as arising from hidden tertiaries.
This authors however, model the full distribution of relative velocities obtained as the sum of Newtonian expectations
and an extra distribution made up of whatever does not conform to Newtonian expectations, which they assign to flybys,
un-bound transient pairs which happen to present projected separations in the sampled range, without there being any
physical association between them. As these authors did not calculate the mass velocity scaling within this extended
tail, it is hard to see if our results are also consistent in this more physical sense, as well as in terms of the
presence of a non-Newtonian high relative velocity extension.

As the authors acknowledge in Pittordis \& Sutherland (2019), the flyby interpretation of the non-Newtonian high relative
velocity remainder to their full distribution once the Newtonian component has been removed, is actually inconsistent with
the expected behaviour of any such flybys. Whilst even in their simulations (as expected under any reasonable modeling of a flyby
population) the fraction of flyby systems increases
as the binary separation range increases, due to the larger cross section for the flyby, their fits to flyby fractions once the
purely Newtonian contribution is subtracted from their observations, require always diminishing flyby fractions as binary
separations increase. It is not only the behaviour of the flyby fraction which is inconsistent, but also the velocity distribution.
As explained previously, if relative velocities are drawn at random from the Gaussian field distribution of velocities, with
a velocity dispersion close to $\sigma=40$ km/s, the expected pairwise velocity will also have a Gaussian distribution, with
a velocity dispersion $\sqrt{2}$ times larger, and hence, for the $\approx$ 1 km/s relative velocities observed for the wide
binaries in question, the flyby distribution would have a very close to constant velocity distribution, and not the elaborate
distributions peaked before Newtonian values and slowly falling thereafter required by these authors once the data are split
between a Newtonian compliant set and an everything else remainder. Further, given the typical interstellar separations of
1 pc, flyby fractions of order 0.5, as required by the authors in question, at binary separations of a few hundredths of a pc,
appear unlikely.

Thus, there is not any evident inconsistency between our current results and those of  Pittordis \& Sutherland (2019),
rather, a clear distinct interpretation of the results. We are merely presenting empirical data which resemble Tully-Fisher
dynamics in the same general acceleration range as relevant for the galactic case (in both separation vs. velocity and mass
vs. velocity scalings), without hazarding any detailed interpretation. Additionally, in our present study we are using GAIA eDR3
data, where typical proper motion errors have gone down by a factor of 2 with respect to the DR2 data used by Pittordis \&
Sutherland (2019), it is therefore entirely possible that a number of systems which they could accommodate within a Newtonian
distribution within the errors, would now have to be assigned to the peculiar flyby high velocity tail.

\section{Discussion}

Whilst it is impossible to be certain that no kinematic contaminants remain in our final sample, the thorough pruning
strategy undertaken, aimed at producing very high quality if small samples, together with the convergence of our results
seen in figures 6b and 7b, strongly suggest an intrinsic origin to the asymptotically flat relative velocity signal obtained
towards large separations in our final sample. The study of the internal kinematics of $s>7,000$ AU binaries is still a very
recent field, where significant observational and theoretic studies are in all probability yet to come. In attempting to reconcile
our still incomplete present data to either Newtonian or MONDian expectations, we find that both paradigms present problems,
as we discuss below.

Within a Newtonian framework, this result is troublesome for a number of reasons. Firstly, we should expect to see a substantial
population of wide binaries with separations larger than $10^{-2}$ pc showing kinematics consistent with Newtonian
expectations, much below the signal we detect. This signal shows no dependence on the separation and has a constant rms value
consistent with 0.5 km/s between $10^{-2}$ and $10^{-1}$ pc. Wide binaries in the above separation range, well within the Newtonian
tidal radius of $\approx$ 0.7 pc, following Keplerian kinematics appear only in very reduced numbers in our GAIA eDR3 sample.

Secondly, it is hard to explain the presence of a clearly defined population of wide binaries in the above separation range with
relative velocities in the plane of the sky having rms values consistent with 0.5 km/s. If these binaries are unbound interlopers,
the lifetime of such systems for the middle of the above range of $5\times 10^{-2}$ pc is a mere
$(5\times 10^{-2}/0.5)\times 10^{6} \approx 10^{5}$ yrs. Thus, given the lifetimes of the stars involved of close to 10 Gyr, the wide
binaries defining the signal we find with rms relative velocity values of close to 0.5 km/s would have to be interpreted as fleeting
transients lasting only of the order of $10^{-5}$ times the lifetimes of the stars involved.

{  Despite the thorough kinematic contamination removal strategy undertaken, we
do expect the presence of a remaining fraction close to 40\% of hidden tertiaries in the separation interval
13.4-100 AU, (resulting in close to one in every 10 of our final binaries, if we assume the constant distribution of
hidden tertiaries per logarithmic interval of Clarke 2020), as per empirical estimates in Tokovinin (2002) and (2010).
For example, a 0.3 $M_{\odot}$ hidden tertiary orbiting at 100 AU from one
member of a Newtonian wide binary composed of two 1 $M_{\odot}$ stars, would modify the relative velocity of the binary
in question, from 0.2 km/s in the absence of the hidden tertiary, to up to about 1 km/s, depending on the
relative orientations of both orbits and ellipticity. However, it appears unlikely that any remaining kinematic contaminants,
hidden tertiaries or flybys, should constitute a dominant kinematic contamination effect, for the reason explained below.

If hidden tertiary contamination were a dominant factor in the non-Newtonian separation region of our final plot, to the
right of 0.009 in figure 7b, and given the empirical absence of a correlation between tertiary fraction and wide binary
separation (e.g. see Tokovinin et al. 2010), such a contaminant would be equally present throughout the Newtonian
region. Towards the small separations, it would clearly be of a very minor relative amplitude, but it would
be equally dominant kinematically just to the left of the transition between both regimes. We have checked
that the clear Newtonian consistent scaling we report in figure 8a for the Newtonian region of figure 7b
is not lost in going to the larger separations within this regime. In fact, for the 0.003-0.009 pc separation
range in figure 7a, we obtain a mass-velocity scaling of $\overline{\Delta V} \propto M_{b}^{0.56 \pm 0.28}$ with a
correlation coefficient of 0.71. The uncertainty in the slope is now somewhat larger than what we obtain for the
much more populated 0.0004-0.009 range in figure 7b of $\overline{\Delta V} \propto M_{b}^{0.52 \pm 0.14}$, but crucially,
remains with Newtonian expectations of 0.5, to well within the confidence intervals.

If kinematic contaminants were dominant for separations above
0.009 pc, they would be equally dominant just to the left of this value, and then we should expect a
constant mass-velocity scaling across the transition. This is not what happens, as to the right of 0.009 pc
the mass-velocity scaling shifts to a much smaller $\overline{\Delta V} \propto M_{b}^{0.24 \pm 0.21}$. The use
of the mass-velocity scalings in the various regions of the plots presented, furnishes an
independent physical diagnostic into the likely causes of the trends seen in the velocity-separation plots.
The clear Newtonian mass-velocity scaling obtained for both the entire separation range showing a Newtonian
velocity-separation scaling in figure 7b and the small 0.003-0.009 region immediately preceding the divergence
from Newtonian expectations, is an indication of the fact that any remaining level of kinematic
contamination is not a dominant contribution, across the entire Newtonian and non-Newtonian regions.}

On the other hand, if they are transients, they are also inconsistent with being chance
wide encounters of pairs of stars being drawn from the average Milky Way stellar velocity distribution at the solar neighbourhood.
As pointed out previously, the distribution of relative velocities of pairs will be a Gaussian having a velocity dispersion of close
to 60 km/s. Given the upper kinematic cut off of 4 km/s we introduced, much smaller than
the pairwise relative velocity dispersion for stars drawn from the overall solar neighbourhood distribution, we should expect a close to
constant distribution of relative velocities for the binaries we study, uniformly distributed between 0 and 4 km/s. Such a
distribution of relative velocities would have a rms value slightly above 2 km/s, much higher than the 0.5 km/s we obtain, and
inconsistent with this value at many times the statistical errors of the extremely high quality GAIA eDR3 catalogue.

Thus, random stars in the
solar neighbourhood which just happen to find themselves within a distance of $5\times 10^{-2}$ pc of each other, would
necessarily show relative velocities much higher than those we find. Further, the average interstellar separation
at the solar neighbourhood of about 1 pc, makes the stars having separations of between $10^{-2}$ and
$10^{-1}$ pc and relative velocities close to 0.5 km/s, inconsistent both with Newtonian bound binaries, and chance passing
encounters within the solar neighbourhood.

If the results we obtained are due to a population of bound wide binaries which display kinematics inconsistent with Newtonian
dynamics, a few first-order estimates become illustrative. Firstly, it is interesting to note that the critical separation beyond which
the observed kinematics cease to follow the Newtonian predictions of Jiang \& Tremaine (2010) is only slightly below, considering the
accuracy allowed by the bin size imposed by the numbers of binaries in our final plot, a value of 0.035 pc. This being the
threshold originally identified in Hernandez et al. (2012a) as the limit separation beyond which a binary system composed of
$1 M_{\odot}$ stars will find itself in the low acceleration $a<a_{0}$ regime, where gravitational anomalies frequently identified
as signalling the presence of dark matter halos in the context of galactic dynamics appear (e.g. Milgrom 1983 or Lelli et al. 2017).

Also, the transition appearing at the above mentioned threshold, is reminiscent of the one between an inner baryonic
dominated regime in galaxies, and the subsequent $a<a_{0}$ flat rotation curve region, in that what we find is precisely such a
Newtonian consistent baryonic small separation regime, which transitions to a constant relative velocity region which would imply,
under a Newtonian scenario, the presence of a singular isothermal dark matter halo. At the length and velocity scales being
probed, such a dominant dark matter halo around all stars would be inconsistent not only with standard structure formation scenarios,
but also with the total mass budget of the Milky Way disk, e.g. Kuijken \& Gilmore (1991). 

Finally, it would appear as an unlikely coincidence, in the absence of a causal connection, that if one extrapolates the baryonic
Tully-Fisher relation between the total baryonic mass of a galaxy and the amplitude of its rotation curve, $V_{TF}=0.35 (M/M_{\odot})^{1/4}$
down to stellar mass scales, e.g. the {  $1.6 M_{\odot}$ average binary masses we infer for the main sequence stars examined in figure 8b},
one should obtain for the corresponding flat rotation amplitude, a value of 0.4 km/s. This last is remarkably close to the rms relative
velocity on the plane of the sky which we obtain for the 'flat rotation' separation region, of $ \approx $ 0.5 km/s.

Within a purelly Newtonian scenario, the possibility of very low luminosity undetected
tertiary or higher hierarchy companions can of course not be ruled out, extremely cold white dwarfs, or even black holes with periods
longer than the GAIA temporal baseline will not produce a significant wobble (Belokurov et al. 2020) of the type the presence of
which we can infer through a decrease in the goodness of fit as evident through increasing RUWE indices. This form of dark matter
would of course be an {\it ad hoc} solution which one would have to introduce coincidentally on crossing the $a<a_{0}$ threshold,
and of just the right amplitude to essentially match the Tully-Fisher extrapolation mentioned above.

On the other hand, our results can not be reconciled with MOND in a straightforward way either. This is because in the most well studied
versions of MOND one expects the appearance of an external field effect, such that if a system is internally in the low acceleration
regime, but embedded within a $a>a_{0}$ or $a \approx a_{0}$ larger system, like the wide binaries we study here which form part of
the solar neighbourhood of the Milky Way, the modifications with respect to Newtonian predictions are expected to be very small,
of the order of a 20\% effect, e.g. Pittordis \& Sutherland (2019). Although MOND is a fundamentally empirical construction, the external
field effect remains mostly a prediction, albeit recent claims of a detection in the specific context of statistical studies of rotation
curves of spiral galaxies, e.g.{  Haghi et al. (2016)} and Chae et al. (2020b).

In the absence of a definitive covariant version of MOND, it is entirely possible that a low velocity limit for a finished theory might
appear where the external field effect is absent (as in e.g. Milgrom 2011), or that such an effect might have mass or scale dependences
different from what results in current well studied versions of MOND. Many modified gravity theories having a low velocity MOND limit
have also been proposed (see e.g. the very incomplete list given in the introduction) where the details, or even existence of the
external field effect, have yet to be explored.

Beyond any theoretical interpretation, the results presented here suggest a 'baryonic Tully-Fisher' phenomenology at stellar
scales, which undoubtedly warrants further investigation, awaiting a definitive confirmation once the current sample can be significantly
enlarged, e.g. once the full GAIA results become available, with the expected significant increase in stars having reliable radial
velocities, or the advent of future generation samples. {  Whether the results presented are a reflection of finely tuned hidden
tertiary distributions, yet unknown astrometric systematics, or are indicative of a low acceleration modification in gravity
is an interesting question which we can hope to answer over the next few years.}

\section*{acknowledgements}

The authors thank an anonymous referee for constructive criticism leading to an enriched and clearer final version. 
The authors are indebted to Mike McCulloch for having originally suggested the collaboration, to Zac Plummer for
his input in large data analysis and to Kareem El-Badry for his assistance implementing the ADQL query used.
Xavier Hernandez acknowledges financial assistance from UNAM DGAPA grant IN106220 and CONACYT ({\it Consejo Nacional
  de Ciencia y Tecnolog{\'i}a}). R.A.M. Cort{\'e}s would like to aknowledge CONACYT for supporting this work through a
scholarship. This work has made use of data from the European Space Agency (ESA) mission {\it Gaia} ({https://www.
  cosmos.esa.int/gaia}), processed by the {\it Gaia} Data Processing and Analysis Consortium (DPAC, {https://www.
  cosmos.esa.int/web/gaia/dpac/consortium}). Funding for the DPAC has been provided by national institutions, in
particular the institutions participating in the {\it Gaia} Multilateral Agreement.

\section*{Data Availability}

The data underlying this article will be shared on reasonable request to the corresponding author.

\section{Appendix:Non-Newtonian binaries}

{  We here present a table giving the GAIA eDR3 identifiers and relevant data for the 128 wide binary pairs showing
relative velocities above Newtonian expectations, those appearing in figure 8a.}

\begin{table*}
\begin{flushleft}
  \caption{Parameters for the binary stars appearing in figure 8b.}
  \label{Table1:example}
  \begin{tabular} {l@{\:}|c@{\:}|c@{\:}|l@{\:}|c@{\:}|c@{\:}|c@{\:}|c@{\:}|c@{\:}|c@{\:}|c@{\:}} 
  \hline
  \hline

GAIA eDR3 ID$_{1}$ & $\varpi_{1}$ & $\sigma \varpi_{1}$ & GAIA eDR3 ID$_{2}$ & $\varpi_{2}$ & $\sigma \varpi_{2}$ & 
$\Delta V_{RA}$ & $\sigma \Delta V_{RA}$ & $\Delta V_{Dec}$ & $\sigma \Delta V_{Dec}$ & s \\

 \hline
 
4725155516009856     & 9.668 &	0.0157  & 4724021644644224  &	9.686 &	0.0166  & 0.426 &       0.331 &	0.232 &	0.102 &	0.0148 \\
36481628308083968    & 8.493 &	0.0197  & 36481593948346240 &	8.521 &	0.0149  & 0.059 &       0.163 &	0.405 &	0.090 &	0.0126 \\
76300510625993344    & 16.631 &	0.0161  & 76300476266255488 & 	16.637 & 0.0163 & 0.385 &       0.074 &	0.431 &	0.110 &	0.0101 \\
98692614681349248    & 29.510 &	0.0224  & 98692339803443328 &	29.537 & 0.0179 & 0.337 &	0.027 &	0.258 &	0.011 &	0.0097 \\ 
104998928046427264   & 16.623 &	0.0176  & 105004842216905344 & 	16.653 & 0.0189	& 0.206 &	0.057 &	0.092 &	0.053 &	0.0324 \\
153741691551129216   & 9.508 &	0.0165  & 153741760270606464 &	9.463 &	0.0164	& 0.361 &	0.052 &	0.102 &	0.048 &	0.0324 \\
371552436752709504   & 17.577 &	0.0162  & 371552402392972032 &	17.557 & 0.0186	& 0.082 &	0.092 &	0.272 &	0.127 &	0.0098 \\
507207363898475008   & 8.405 &	0.0112  & 507207329538738432 & 	8.413 &	0.0136	& 0.353 &       0.065 &	0.059 &	0.030 &	0.0100 \\
644549800855341184   & 10.808 &	0.0177  & 644549869574817792 &	10.814 & 0.0173	& 0.265 &	0.087 &	0.082 &	0.054 &	0.0114 \\
680661782802091392   & 9.930 &	0.0192  & 680662573076074752 &	9.943  & 0.0195	& 0.183 &	0.030 &	0.159 &	0.043 &	0.0677 \\
758958211973432704   & 14.957 &	0.0186  & 758958929232265472 &	14.981 & 0.0213	& 0.939 &	0.077 &	0.427 &	0.045 &	0.0280 \\ 
796311542548505984   & 10.197 &	0.0180  & 796311954865365888 &	10.275 & 0.0164	& 0.811 &	0.259 &	0.246 &	0.047 &	0.0310 \\
922601585552127104   & 15.912 &	0.0196	& 922595430863222144 &	15.878 & 0.0143	& 0.721 &	0.084 &	1.147 &	0.016 &	0.0495 \\
992789150829982848   & 10.962 &	0.0161	& 992789872384490496 &	10.972 & 0.0181	& 0.219 &	0.069 &	0.494 & 0.067 &	0.0122 \\
1021917447232306816  & 11.292 &	0.0132	& 1021917447232306944 &	11.315 & 0.0127	& 0.067 &	0.050 &	0.548 &	0.012 &	0.0094 \\
1023887978228642176  & 10.788 &	0.0182	& 1023888012588380032 &	10.784 & 0.0191	& 0.295 &	0.175 &	0.465 &	0.094 &	0.0094 \\
1055226293001395840  & 10.137 &	0.0126	& 1055226219986947200 & 10.101 & 0.0125	& 0.775 &	0.088 &	0.611 & 0.039 &	0.0267 \\
1065304244783926400  & 9.242 &	0.0109	& 1065304068689257856 &	9.258 &	0.0109	& 0.001 &	0.117 &	0.136 &	0.043 &	0.0388 \\
1112877875238798592  & 7.727 &	0.0189	& 1112877913895749888 &	7.693 &	0.0158	& 0.275 &	0.101 &	0.411 & 0.065 &	0.0190 \\ 
1129061930485416960  & 10.839 &	0.0121	& 1129061797342364928 &	10.828 & 0.0117	& 0.089 &       0.115 &	0.901 &	0.026 &	0.0548 \\
1142787168495168000  & 9.477 &	0.0108	& 1142786996696476288 &	9.485 &	0.0117	& 0.100 &	0.059 &	0.224 &	0.031 &	0.0183 \\
1227417304334766208  & 8.263 &	0.0148	& 1227416514060783232 &	8.240 &	0.0142	& 0.180 &	0.092 &	0.422 &	0.040 &	0.0214 \\
1346694146783637504  & 11.163 &	0.0092	& 1346694112423491712 & 11.156 & 0.0116	& 0.764 &	0.017 &	0.287 &	0.039 &	0.0111 \\
1348285896022947584  & 20.547 &	0.0162	& 1348286651937191040 &	20.554 & 0.0108	& 0.152 &	0.031 &	0.021 & 0.010 &	0.0321 \\
1509241238549271808  & 8.495 &	0.0127	& 1509241238549292800 &	8.479 &	0.0151	& 0.153 &	0.039 &	0.051 &	0.033 &	0.0246 \\
1549521125478552064  & 20.469 &	0.0128	& 1549520949383005568 &	20.464 & 0.0153	& 0.076 &	0.013 &	0.045 &	0.027 &	0.0172 \\
1580278623234613248  & 7.988 &	0.0097	& 1580278623234613376 &	7.979 &	0.0101	& 0.062 &	0.046 &	0.053 &	0.051 &	0.0094 \\
1584402341594679808  & 13.138 &	0.0133	& 1584402479033634048 &	13.084 & 0.0153	& 0.426 &	0.028 &	0.236 &	0.041 &	0.0281 \\
1585800851666242688  & 8.891 &	0.0115	& 1585800782946765312 & 8.895 &	0.0112	& 0.220 &	0.037 &	0.362 &	0.075 &	0.0228 \\
1586977844504488576  & 29.624 &	0.0135	& 1586977737129182848 &	29.610 & 0.0115	& 0.252 &	0.009 &	0.188 &	0.018 &	0.0103 \\
1661173816859019264  & 10.701 &	0.0110	& 1661174023017450112 & 10.716 & 0.0110	& 0.675 &	0.125 &	0.273 & 0.023 &	0.0394 \\
1696726250465683968  & 8.822 &	0.0130	& 1696726319185160960 &	8.777 &	0.0124	& 0.897 &	0.066 &	0.262 &	0.053 &	0.0193 \\
1700110684694632832  & 12.843 &	0.0113	& 1700112157867455232 &	12.863 & 0.0105	& 0.151 &	0.057 & 0.197 &	0.035 &	0.0095 \\
1709320297168157824  & 9.003 &	0.0109	& 1709320292871977856 &	9.001 &	0.0105	& 0.524 &	0.029 &	0.009 &	0.057 &	0.0124 \\
1719835231806217472  & 10.033 &	0.0116	& 1719835407900844544 &	10.042 & 0.0104	& 0.102 &	0.098 &	0.278 &	0.033 &	0.0411 \\
1760471948915107200  & 13.272 &	0.0162	& 1760477618271932672 &	13.242 & 0.0189	& 0.049 &	0.024 &	0.024 &	0.050 &	0.0207 \\
1778929480673414784  & 11.025 &	0.0195	& 1778930240883745536 &	11.069 & 0.0146	& 0.298 &	0.156 &	0.888 &	0.036 &	0.0302 \\
1952145206786538112  & 9.554 &	0.0124	& 1952145172426798208 &	9.532  & 0.0136	& 0.179 &	0.110 & 0.032 &	0.087 &	0.0125 \\
2140767560397457536  & 13.707 &	0.0109	& 2140767319879292160 &	13.717 & 0.0101	& 0.301 &	0.038 &	0.076 &	0.032 &	0.0352 \\
2201834466266276864  & 10.213 &	0.0101	& 2201834470572701952 & 10.214 & 0.0101	& 0.024 &	0.030 &	0.036 &	0.062 &	0.0182 \\
2211667368695978240  & 15.087 &	0.0117	& 2211667368695552896 &	15.068 & 0.0103	& 0.232 &	0.019 &	0.018 &	0.030 &	0.0106 \\
2233231540491361408  & 15.150 &	0.0111	& 2233231330034854656 &	15.173 & 0.0102	& 0.455 &	0.010 &	0.116 & 0.022 &	0.0129 \\
2260460155678348160  & 11.351 &	0.0107	& 2260460224397823488 &	11.337 & 0.0126	& 0.476 &	0.020 &	0.152 &	0.047 &	0.0127 \\

\hline 
  \end{tabular}
  \end{flushleft}
\end{table*}

\begin{table*}
\begin{flushleft}
  \contcaption{Parameters for the binary stars appearing in figure 8b.}
  \label{Table1:continued}
  \begin{tabular} {l@{\:}|c@{\:}|c@{\:}|l@{\:}|c@{\:}|c@{\:}|c@{\:}|c@{\:}|c@{\:}|c@{\:}|c@{\:}} 
  \hline
  \hline

GAIA eDR3 ID$_{1}$ & $\varpi_{1}$ & $\sigma \varpi_{1}$ & GAIA eDR3 ID$_{2}$ & $\varpi_{2}$ & $\sigma \varpi_{2}$ & 
$\Delta V_{RA}$ & $\sigma \Delta V_{RA}$ & $\Delta V_{Dec}$ & $\sigma \Delta V_{Dec}$ & s \\

 \hline

2337837908524395776  & 8.443 &	0.0196	& 2337837908524639744 &	8.460 &	0.0195	& 0.010 &	0.123 &	0.521 &	0.034 &	0.0152 \\
2386641518829641984  & 13.841 &	0.0196	& 2386641312671208960 &	13.837 & 0.0168	& 0.460 &	0.017 &	1.023 &	0.045 &	0.0552 \\
2436430016675388160  & 7.951 &	0.0166	& 2436430119754603008 &	8.028 &	0.0175	& 1.006 &	0.192 &	0.625 & 0.094 &	0.0517 \\
2514529598906714880  & 13.704 &	0.0230	& 2514541414361083648 &	13.776 & 0.0165	& 0.375 &	0.064 &	0.371 &	0.076 &	0.0237 \\
2522327232292290688  & 18.212 &	0.0156	& 2522325548665111424 & 18.233 & 0.0195	& 0.271 &	0.021 &	0.233 &	0.036 &	0.0169 \\
2592845200813026688  & 8.210 &	0.0128	& 2592845200813028992 &	8.183 &	0.0141	& 0.194 &	0.218 &	0.087 &	0.193 &	0.0187 \\
2614582163441455616  & 8.876 &	0.0195	& 2614582064657951104 &	8.877 &	0.0190	& 0.234 &	0.200 &	0.338 &	0.179 &	0.0201 \\
2663015998537909760  & 20.132 &	0.0172	& 2663015994242895488 &	20.130 & 0.0154	& 0.401 &	0.030 &	0.607 &	0.037 &	0.0118 \\
2840600492363295616  & 14.914 &	0.0143	& 2840599633369837312 &	14.951 & 0.0137	& 0.104 &	0.032 &	0.063 &	0.035 &	0.0425 \\
2989732545838913536  & 7.967 &	0.0156	& 2989735565198378368 &	7.958 &	0.0144	& 0.182 &	0.071 &	0.029 &	0.036 &	0.0563 \\
3046204180298475264  & 11.459 &	0.0147	& 3046204150242468480 &	11.472 & 0.0156	& 0.045 &	0.092 &	0.143 &	0.046 &	0.0297 \\
3072944474884690688  & 10.809 &	0.0160	& 3072920942760216064 &	10.825 & 0.0154	& 0.076 &	0.024 &	0.078 &	0.022 &	0.0468 \\
3076534861386561536  & 14.437 &	0.0163	& 3076535243639085440 &	14.433 & 0.0169	& 0.541 &	0.112 &	0.484 &	0.072 &	0.0133 \\
3158926322836178816  & 12.818 &	0.0138	& 3158878734598549376 &	12.814 & 0.0133	& 0.043 &	0.031 &	0.201 &	0.026 &	0.0532 \\
3194720889516362880  & 8.318 &	0.0165	& 3194720786437148032 &	8.315 &	0.0146	& 0.295 &	0.153 &	0.206 &	0.048 &	0.0126 \\
3199303963218457088  & 9.007 &	0.0158	& 3199303963218456832 &	8.948 &	0.0156	& 0.213 &	0.035 &	0.268 &	0.047 &	0.0227 \\
3286839549942183168  & 8.441 &	0.0153	& 3286839932195653888 &	8.452 &	0.0155	& 0.115 &	0.125 &	0.389 &	0.020 &	0.0279 \\
3371529368651025664  & 17.554 &	0.0185	& 3371529467432172672 &	17.581 & 0.0184	& 0.237 &	0.053 &	0.110 &	0.020 &	0.0154 \\
3402090259984528768  & 13.321 &	0.0192	& 3402090466142958464 &	13.300 & 0.0243	& 0.298 &	0.047 & 0.279 &	0.036 &	0.0162 \\
3431938839582893568  & 7.988 &	0.0197	& 3431938766565511424 &	8.053 &	0.0135	& 0.525 &	0.036 &	0.408 &	0.086 &	0.0305 \\
3451267120128948992  & 24.064 &	0.0196	& 3451266742171824640 &	24.044 & 0.0225	& 0.487 &	0.015 &	0.205 &	0.010 &	0.0121 \\
3497346209336507520  & 8.002 &	0.0260	& 3497346205040949248 &	8.085 &	0.0189	& 0.825 &	0.312 &	1.775 &	0.348 &	0.0160 \\
3538247870092026624  & 13.531 &	0.0155	& 3538249588079338112 &	13.600 & 0.0233	& 0.038 &	0.092 &	0.006 &	0.024 &	0.0515 \\
3550081879381593728  & 29.831 &	0.0262	& 3550084490721711872 &	29.799 & 0.0178	& 0.045 &	0.036 &	0.215 &	0.014 &	0.0382 \\
3557719293306114560  & 17.682 &	0.0156	& 3557719293306114944 &	17.717 & 0.0160	& 0.113 &	0.077 &	0.141 &	0.034 &	0.0092 \\
3731743568479226880  & 8.399 &	0.0181	& 3731742812564982912 &	8.393 &	0.0162	& 0.375 &	0.253 &	0.038 &	0.224 &	0.0136 \\
3792739899447945216  & 9.297 &	0.0145	& 3792740006822516608 &	9.240 &	0.0150	& 0.206 &	0.078 &	0.762 &	0.163 &	0.0128 \\
3836090722352641536  & 8.882 &	0.0221	& 3836090726648115968 &	8.856 &	0.0189	& 0.082 &	0.117 &	0.023 &	0.221 &	0.0147 \\
3907643060734192896  & 17.653 &	0.0190	& 3907643060733826432 &	17.696 & 0.0191	& 0.040 &	0.166 &	0.292 &	0.020 &	0.0115 \\
3975129194660883328  & 25.376 &	0.0270	& 3975223065466473216 &	25.375 & 0.0240	& 0.005 &	0.181 &	0.197 &	0.014 &	0.0140 \\
4218533748765026560  & 12.139 &	0.0149	& 4218533959216594176 &	12.141 & 0.0142	& 0.300 &	0.105 &	0.408 &	0.099 &	0.0116 \\
4395523033138822656  & 9.178 &	0.0197	& 4395522998779760128 &	9.204 &	0.0190	& 0.072 &	0.028 &	0.414 &	0.086 &	0.0102 \\
4401991146507698944  & 	7.787 &	0.0139	& 4401990394889506304 &	7.818 &	0.0117	& 0.317 &	0.061 &	0.288 &	0.035 &	0.0144 \\
4404823011724525184  & 	8.574 &	0.0191	& 4404823011724524416 &	8.586 &	0.0167	& 0.010 &       0.041 &	0.296 &	0.106 &	0.0125 \\
4407544268641139712  & 	9.432 &	0.0169	& 4407541313703639424 &	9.457 &	0.0170	& 0.517 &	0.045 &	0.213 &	0.059 &	0.0107 \\
4430185068482324864  & 	9.967 &	0.0140	& 4430185034123000960 &	9.986 &	0.0152	& 0.240 &	0.066 &	0.221 &	0.016 &	0.0214 \\
4474801257476653952  & 	8.125 &	0.0131	& 4474801218817508864 &	8.160 &	0.0121	& 0.055 &	0.023 & 0.104 &	0.043 &	0.0112 \\
4545496243069399808  &	14.315 & 0.0146	& 4545496075567671808 &	14.311 & 0.0152	& 0.031 &	0.059 &	0.805 &	0.047 &	0.0148 \\
4577272270853565952  &	10.531 & 0.0107	& 4577271961615916032 &	10.525 & 0.0130	& 0.175 &	0.063 &	0.395 &	0.033 &	0.0203 \\
4584470739116877056  & 	8.129 &	0.0133	& 4584470704757140096 &	8.151 &	0.0112	& 0.266 &	0.049 &	0.005 &	0.065 &	0.0435 \\
4628920897552913792  & 	9.532 & 0.0106	& 4628920970569782400 &	9.495 &	0.0099	& 0.330 &	0.085 &	0.085 &	0.045 &	0.0396 \\
4629399567364049280  & 	8.620 &	0.0108	& 4629399567364049664 &	8.597 &	0.0105	& 0.263 &	0.240 &	0.439 &	0.211 &	0.0124 \\
4701607897573989248  & 	10.438 & 0.0135	& 4701607893278187648 &	10.446 & 0.0112	& 0.589 &	0.078 &	0.127 & 0.060 &	0.0156 \\
4729517591496033920  & 	9.250 &	0.0096	& 4729517591496033152 &	9.226 &	0.0110	& 0.154 &	0.019 &	0.200 &	0.023 &	0.0098 \\

\hline 
  \end{tabular}
  \end{flushleft}
\end{table*}

\begin{table*}
\begin{flushleft}
  \contcaption{Parameters for the binary stars appearing in figure 8b.}
  \label{Table1:continued}
  \begin{tabular} {l@{\:}|c@{\:}|c@{\:}|l@{\:}|c@{\:}|c@{\:}|c@{\:}|c@{\:}|c@{\:}|c@{\:}|c@{\:}} 
  \hline
  \hline

GAIA eDR3 ID$_{1}$ & $\varpi_{1}$ & $\sigma \varpi_{1}$ & GAIA eDR3 ID$_{2}$ & $\varpi_{2}$ & $\sigma \varpi_{2}$ & 
$\Delta V_{RA}$ & $\sigma \Delta V_{RA}$ & $\Delta V_{Dec}$ & $\sigma \Delta V_{Dec}$ & s \\

 \hline

4810651860180442624  &	10.972 & 0.0091	& 4810651928899919616 &	11.014 & 0.0098	& 0.170 &	0.010 &	0.186 &	0.088 &	0.0091 \\
4813061852229522432  &	13.247 & 0.0101	& 4813061676134636928 &	13.273 & 0.0099	& 0.390 &	0.026 &	0.023 &	0.063 &	0.0154 \\
4819741759028427904  &	8.253 &	0.0122	& 4819741763325357056 &	8.275 &	0.0136	& 1.958 &	0.027 &	0.334 &	0.023 &	0.0093 \\
4822945740208994176  &	9.424 &	0.0118	& 4822945705849256192 &	9.423 &	0.0126	& 0.292 &	0.052 &	0.151 &	0.034 &	0.0118 \\
4876470412926321280  &	9.184 &	0.0145	& 4876443990287014912 &	9.211 &	0.0115	& 0.146 &	0.073 &	0.025 &	0.053 &	0.0234 \\
4879098280075566208  & 	13.442 & 0.0108	& 4879098383154780672 &	13.461 & 0.0119	& 0.182 &	0.038 &	0.326 &	0.126 &	0.0154 \\
4899504459972848512  & 	13.027 & 0.0153	& 4899498069061515008 &	13.019 & 0.0155	& 0.117 &	0.036 &	0.002 &	0.169 &	0.0928 \\
4902812787380558208  &	7.760 &	0.0145	& 4902812787380558976 &	7.781 &	0.0117	& 0.486 &	0.024 &	0.248 &	0.213 &	0.0096 \\
4915145322115519104  &	8.021 &	0.0102	& 4915145322115519232 &	7.980 &	0.0103	& 0.008 &	0.021 &	0.371 &	0.019 &	0.0096 \\
4981101348174544512  &	8.474 &	0.0116	& 4981101348174544768 &	8.433 &	0.0127	& 0.754 &	0.047 &	0.346 &	0.107 &	0.0201 \\
5036326759219158272  & 	9.258 &	0.0140	& 5035575758417810688 &	9.248 &	0.0142	& 0.025 &	0.033 &	0.120 &	0.061 &	0.0438 \\
5099953023216735872  &	9.557 &	0.0157	& 5099953057577400704 &	9.553 &	0.0146	& 0.076 &	0.236 &	0.273 &	0.071 &	0.0314 \\
5114547700047886976  &	11.807 & 0.0155	& 5114544745110388352 &	11.793 & 0.0145	& 0.058 &	0.175 &	0.098 &	0.087 &	0.0178 \\
5179618511869510272  &	8.477 &	 0.0155	& 5179618752387678592 &	8.463 &	0.0170	& 1.756 &	0.063 &	0.359 &	0.203 &	0.0659 \\
5181911238426924288  &	8.353 &	0.0153	& 5181911234131665920 &	8.409 &	0.0162	& 0.342 &	0.051 &	0.226 &	0.035 &	0.0121 \\
5269452631947873152  & 	8.166 &	0.0091	& 5269452636245616128 &	8.130 &	0.0102	& 0.269 &	0.055 &	0.118 &	0.155 &	0.0169 \\
5796674577807917184  &	20.093 & 0.0124	& 5796675471162119040 &	20.090 & 0.0139	& 0.158 &	0.007 &	0.141 &	0.016 &	0.0282 \\
5800969407655752192  &	8.263 &	 0.0115	& 5800969235854202496 &	8.289 &	0.0107	& 0.053 &	0.031 &	0.071 &	0.017 &	0.0402 \\
5891531404470141824  & 	38.998 & 0.0228	& 5891544873500692608 &	39.050 & 0.0145	& 0.151 &	0.050 &	0.035 &	0.043 &	0.0602 \\
6078554071716755456  &	8.231 &	0.0171	& 6078554140436217728 &	8.269 &	0.0171	& 0.339 &	0.195 &	0.020 &	0.015 &	0.0871 \\
6193279279612173952  &	33.480 & 0.0312	& 6193280031230266752 &	33.433 & 0.0227	& 0.053 &	0.085 &	0.292 &	0.020 &	0.0463 \\
6206044369095120512  &	8.295 &	0.0190	& 6206045125009768192 &	8.356 &	0.0273	& 0.007 &	0.172 &	0.520 &	0.153 &	0.0157 \\
6238116298643720192  &	7.778 &	0.0174	& 6238116298643719552 &	7.789 &	0.0155	& 0.182 &	0.057 &	0.128 &	0.062 &	0.0215 \\
6249410860041044480  & 	9.608 &	0.0202	& 6249411203638655232 &	9.647 &	0.0216	& 0.050 &	0.136 &	0.299 &	0.165 &	0.0442 \\
6378374125048999424  &	9.181 &	0.0138	& 6378374052033393024 &	9.211 &	0.0124	& 0.555 &	0.075 &	0.029 &	0.031 &	0.0140 \\
6398232816875801856  &	8.014 &	0.0142	& 6398232778219135744 &	8.022 &	0.0104	& 0.575 &	0.150 &	0.151 &	0.124 &	0.0240 \\
6447718261829058816  &	10.914 & 0.0164	& 6447718467987494784 &	10.918 & 0.0171	& 0.056 &	0.173 &	0.248 &	0.054 &	0.0351 \\
6458626310529028352  &	13.330 & 0.0129	& 6458626271874014208 &	13.315 & 0.0130	& 0.027 &	0.060 &	0.009 &	0.013 &	0.0292 \\
6464392012066637440  &	9.391 &	0.0146	& 6464391977706899712 &	9.385 &	0.0163	& 0.168 &	0.024 & 0.153 &	0.020 &	0.0390 \\
6562582828536563328  &	10.555 & 0.0142	& 6562581969543104640 &	10.554 & 0.0120	& 0.020 &	0.179 &	0.046 &	0.053 &	0.0623 \\
6586315275923449856  &	8.471 &	0.0211	& 6586315207203972224 &	8.504 &	0.0215	& 0.136 &	0.190 &	0.285 &	0.085 &	0.0182 \\
6615293145390228352  &	8.513 &	0.0189	& 6615291667921479552 &	8.535 &	0.0178	& 1.031 &	0.055 &	0.867 &	0.028 &	0.0125 \\
6620898657532005120  &	29.995 & 0.0148	& 6620893645304242176 &	29.959 & 0.0200	& 0.059 &	0.067 &	0.222 &	0.044 &	0.0262 \\ 
6639645128923820160  &	10.611 & 0.0135	& 6639645128923818624 &	10.626 & 0.0145	& 0.007 &	0.011 &	0.028 &	0.012 &	0.0114 \\
6658696396962053120  &	8.362 &	0.0180	& 6658696401257998464 &	8.365 &	0.0159	& 0.223 &	0.045 &	0.256 &	0.018 &	0.0179 \\
6681883211702702208  &	8.539 &	0.0194	& 6681883039904009856 &	8.595 &	0.0174	& 0.329 &	0.120 &	0.550 &	0.042 &	0.0203 \\
6695911949280082432  &	9.200 &	0.0151	& 6695911880560604672 &	9.193 &	0.0134	& 0.424 &	0.082 &	0.151 &	0.032 &	0.0102 \\
6768864992462040832  &	14.645 & 0.0135	& 6768865714016553344 &	14.623 & 0.0144	& 0.385 &	0.051 &	0.037 &	0.036 &	0.0261 \\
6812424241536340864  &	21.290 & 0.0367	& 6812422832786523776 &	21.189 & 0.0178	& 2.871 &	0.134 &	0.590 &	0.013 &	0.0281 \\
6840365718216434816  &	14.041 & 0.0222	& 6840365615137220096 &	14.038 & 0.0158	& 0.319 &	0.070 &	0.083 &	0.066 &	0.0219 \\
6859531958238411904  &	12.022 & 0.0180	& 6859531752079979776 &	11.999 & 0.0233	& 0.011 &	0.025 &	0.287 &	0.061 &	0.0458 \\

 \hline 
 \end{tabular}
  \contcaption{The first three columns give GAIA eDR3 identifiers, parallax and parallax errors for the primaries, while the following three
  columns give the corresponding information for the secondaries. Columns 7-10 give relative velocities for the pair in R.A.,
  the error in this quantity, relative velocities for the pair in Dec. and the error for this quantity. The final column shows
  the projected separation on the pane of the sky for each binary pair. Units for parallaxes, velocities and separations are
  mas, km/s and pc, respectively.}

\end{flushleft}
\end{table*}

\end{document}